\documentclass[bibyear]{aa}  
\usepackage{graphicx,amssymb}
\usepackage[varg]{txfonts}
\begin{document}
   \title{SALT long-slit spectroscopy of CTS C30.10: two-component Mg II line \thanks{based on observations made with the Southern African Large Telescope (SALT) under program 2012-2-POL-003 and 2013-1-POL-RSA-002  (PI: B. Czerny)}}

   \author{J. Modzelewska\inst{1}\and B. Czerny\inst{1}  \and K. Hryniewicz\inst{1,3} \and M. Bilicki\inst{2} \and  M. Krupa\inst{4} \and A. \' Swi\c eto\' n\inst{4} \and W. Pych\inst{1} \and A. Udalski\inst{5}  \and T. P. Adhikari\inst{1} \and F. Petrogalli\inst{1}  
          }
   \institute{Nicolaus Copernicus Astronomical Center, Bartycka 18, 00-716 Warsaw, Poland 
\and
Astrophysics, Cosmology and Gravity Centre, Department of Astronomy, University of Cape Town, Rondebosch, South Africa
\and
ISDC Data Centre for Astrophysics, Observatoire de Geneve, Universite de Geneve, Chemin d’Ecogia 16, 1290 Versoix,  Switzerland
\and
 Astronomical Observatory of the Jagiellonian University, Orla 171, 30-244 Cracow, Poland
\and 
Warsaw University Observatory, Al. Ujazdowskie 4, 00-478 Warszawa, Poland
             }

   \date{Received ............; accepted ..............}

 
  \abstract
   {Quasars can be used as a complementary tool to SN Ia to probe the distribution of dark energy in the Universe by measuring the time delay of the emission line with respect to the continuum. The understanding of the Mg II emission line structure is important for cosmological application and for the black hole mass measurements of intermediate redshift quasars.}
   {The knowledge of the shape of Mg II line and its variability allows to determine which part of the line should be used to measure the time delay and the black hole mass.  We thus aim at determination of the structure and the variability of the Mg II line as well as of the underlying Fe II pseudo-continuum.}
   {We performed five spectroscopic observations of a quasar CTS C30.10 (z = 0.9000) with the SALT telescope between December 2012 and March 2014,
 and we studied the variations of the spectral shape in the 2700 \AA~-2900 \AA~ rest frame.}
   {We showed that the Mg II line in this source consists of two kinematic components, which makes the source representative of type B quasars. Both components were well modeled with a Lorentzian shape. Both components vary in a similar way. The Fe II contribution seems to be related only to the first (blue) Mg II component. Broad band spectral fitting rather favor the use of the whole line profile. The contribution of the Narrow Line Region to Mg II is very low, below 2\%. The Mg II variability is lower than the variability of the continuum, which is consistent with the simple reprocessing scenario. The variability level of CTS C30.10 and the measurement accuracy of the line and continuum is high enough to expect that further monitoring will allow to measure the time delay between the Mg II line and continuum.}
   {}

\authorrunning{Modzelewska et al.}
\titlerunning{SALT Long-slit Spectroscopy of CTS C30.10}

   \maketitle
%

\section{Introduction}

Quasars represent high luminosity tail of active galactic nuclei (AGN). They are numerous, and they can be easily detected at cosmological distances, so they can be used to investigate the properties of the Universe. They are used to probe the intergalactic medium (e.g. Sargent 1977; Bordoloi et al. 2014), they
provide information on the massive black hole growth (e.g. Kelly et al. 2010), and they are recently proposed as promising tracers of the expansion of the Universe (Watson et al. 2011, Czerny et al. 2013, Marziani \& Sulentic 2013, 2014; Wang et al. 2013; Hoenig 2014; Yoshii et al.2014). The last two aspects mostly relay on the presence of the broad emission lines in the quasar spectra.  The broad line region (BLR) is not resolved but the shapes of the lines, and in particular the variability of the lines and of the
continuum are keys to understand the quasar structure. This is interesting by itself, and essential for the follow up applications. 

The systematic studies of the continuum and line variability has been extensively done for nearby AGN since many years (Clavel et al. 1991; Peterson 1993; Reichert et al. 1994; Wandel et al. 1999; Kaspi et al. 2000; Peterson et al. 2004; Metzroth et al. 2006; Bentz et al. 2013). Most studies were done for H$\beta$  line since this line is easily measured in the optical band for sources with redshift below $\sim 0.3$, the line belongs to the family of the Low Ionization Lines (Collin-Souffrin et al. 1988) which do not show considerable shifts with respect to the host galaxy, and finally the nearby narrow [OIII] forbidden line helps to calibrate the spectra. Mg II and CIV lines were followed when there was
UV data available, but for high redshift quasars they are the only alternative for the studies in the optical band. Mg II is suitable for sources with redshifts between 0.4 and 1.5, and CIV for quasars with redshift above 1.7. In the present paper we concentrate on Mg II line.

H$\beta$ line was found to be generally a good indicator of the black hole mass, and the emitting material seems indeed roughly in Keplerian
motion. Several authors (e.g. Kong et al. (2006); Shen et al. 2008;  Vestergaard \& Osmer 2009) concluded that Mg II also gives quite satisfactory results for the black hole mass measurement. Vanden Berk et al. (2001) composite quasar spectrum from SDSS also indicated no shifts of Mg II with respect to the expected position. However, with the increasing pressure on the accuracy, several problems started to appear. Wang et al. (2009) showed that  the Full Width at the Half Maximum (FWHM) of Mg II in many sources is lower by $\sim 20$ \% than FWHM of H$\beta$. Relations between the two lines can be thus found but with a considerable scatter (Shen \& Liu 2012; Trakhtenbrot \& Netzer 2012). Marziani et al. (2013ab)  analyzed 680 quasars from Sloan Digital Sly Survey (SDSS) with both lines visible in their spectra and concluded that, overall, Mg II is a good black hole mass indicator if carefully used. They analyzed the results following the quasar division of Sulentic et al. (2007) into two classes. For type A sources, with FWHM of H$\beta$ below 4000 km s$^{-1}$, the Mg II line is simply narrower than H$\beta$ by 20 \%. The line is well fit by a single Lorentzian. We found the same single Lorentzian shape  in excellent SALT spectrum of LBQS  2113-4538 which also belongs to this class. However, for type B sources (with FWHM of H$\beta$ above 4000 kms $^{-1}$ the Mg II line consists of two components, with a nearly unshifted broad component and redshifted very broad component (Marziani et al. 2009, Marziani et al. 2013b). Marziani et al. (2013b) suggest that only the broad component should be used for black hole mass determination.

The results of Marziani et al. (2013a,b) were based on the analysis of the composite spectra since a single SDSS spectrum has too low signal-to-noise ratio to perform a multicomponent fit reliably. With SALT data, we can do such a decomposition for bright sources. The object CTS C30.10 discussed in this paper is of type B class and requires such complex approach.


\section{Observations}

CTS C30.10 is one of the quasars found in the Calan-Tololo Survey aimed to identify new bright quasars in the southern part of the sky. 
The source is located at RA = 04h47m19.9s, DEC = -45d37m38s.
The quasar nature of the source was confirmed using the slit spectroscopy (Maza et al. 1993). The source is relatively bright for an intermediate redshift quasar 
(z = 0.910, V = 17.2, as given in NED\footnote{NASA/IPAC Extragalactic Database (NED) is operated by the Jet Propulsion 
Laboratory, California Institute of Technology}) so  we selected it for a detailed study of the Mg II line, together with LBQS 2113-4538 (Hryniewicz et al. 2014).

\subsection{Spectroscopy}

We observed the quasar CTS C30.10 with the use of the Robert Stobie Spectrograph 
(RSS; Burgh et al. 2003, Kobulnicky et al. 2003; Smith et al. 2006) on the Southern
African Large Telescope (SALT) in the service mode. The observations were made on the nights of 
Dec 6/7 in 2012 and January 21/22,  March 20/21 and  August 4/5 2013,  and March 5/6 in 2014. Every observation consisted of two 739  second exposures in a long slit mode, with the slit width of 2", an exposure of the calibration lamp, and several flat-field images. As for
LBQS  2113-4538, we used   RSS PG1300 grating, corresponding to the spectral resolution of  $R=1047$ at 5500 \AA. We used the blue PC04600 filter for the order blocking.  
 
The nights were photometric, without intervening clouds. Observations 1,2, 3 and 5 were performed in grey moon conditions, and observation 4 in the dark moon conditions. The seeing was good in observation 4 ($\sim 1.4$) and somewhat worse in the previous three data sets ($\sim 1.5 - 2.4$) as well as in the last observation ($\sim 1.9$).

The basic data reduction was done  by SALT staff using a semi-automated pipeline from the SALT PyRAF package\footnote{http://pysalt.salt.ac.za} (see Crawford et al. 2010). Further steps, including flat-field correction were done by us with the help of IRAF package\footnote{IRAF is distributed by the National 
Optical Astronomy Observatories, which are operated
by the Association of Universities for Research in Astronomy, Inc., under cooperative agreement
with the NSF.}. Two consecutive exposures were combined to increase the signal-to-noise ratio and to remove the cosmic ray effects.
After proper wavelength calibration, we extracted one-dimensional spectra using noao.twodspec package within IRAF.

Since SALT telescope has considerable problems with vigneting that affects  the broad band spectral shape, we used SALT observations of the spectroscopic standards. A number of stars with well calibrated spectra available at ESO were observed by SALT with similar telescope setup as used in our quasar observations. We finally selected the star LTT1020, which was observed with SALT on Oct. 3, 2012, with the instrumental setup PG1300/23.375 (spectrum P201210130322). This is a relatively bright G-type star (V = 11.52) not showing any significant absorption lines in the 5000 - 5600 \AA ~band. The ESO spectrum for this star was downloaded from the ESO website\footnote{ftp://ftp.eso.org/pub/stecf/standards/ctiostan/}. 
We obtained the ratio of the star spectrum from SALT to the calibrated star spectrum from ESO in the 5000 - 5700 \AA~  wavelength range. In next step we normalized this ratio to 1 in the middle of the interesting wavelength range, and  we fit this ratio with a quadratic function.  The derived analytical formula was treated as a telescope response for the five SALT spectra of quasar CTS C30.10. The same correction was used for all spectra. The method allowed us to obtain the correct shape of the spectra but not the absolute calibration. However, this already allows for the determination of the emission lines in the individual spectra.  At that stage the spectra were ready for further analysis. We neglected the instrumental broadening since we showed before (Hryniewicz et al. 2014) that it is unimportant for SALT quasar emission lines.

The effect of the Galactic extinction was removed although the extinction in the direction of CTS C30.10 is
very low  ($A_{\lambda} = 0.038, 0.029$, and 0.023 in the B,V and R bands; Schlafly \& Finkbeiner 2011 
after NED). 

We neglected the intrinsic absorption as there is no clear signature of such an extinction
in the spectra. We also neglected a possible host galaxy contribution as it is not likely 
to be important at such short wavelengths.

\subsection{Photometry}

\begin{figure}
   \centering
   \includegraphics[width=0.95\hsize]{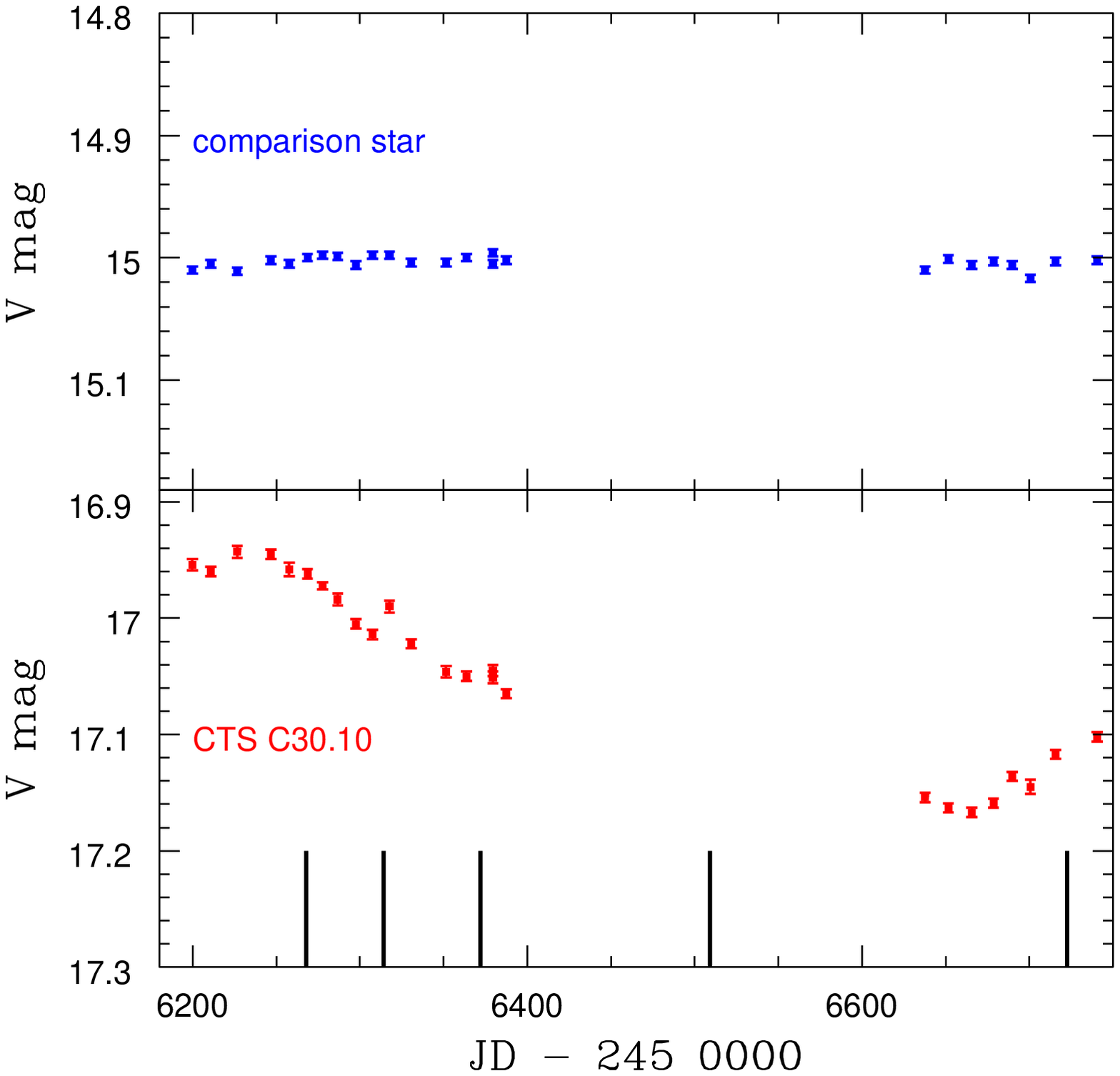}
   \includegraphics[width=0.95\hsize]{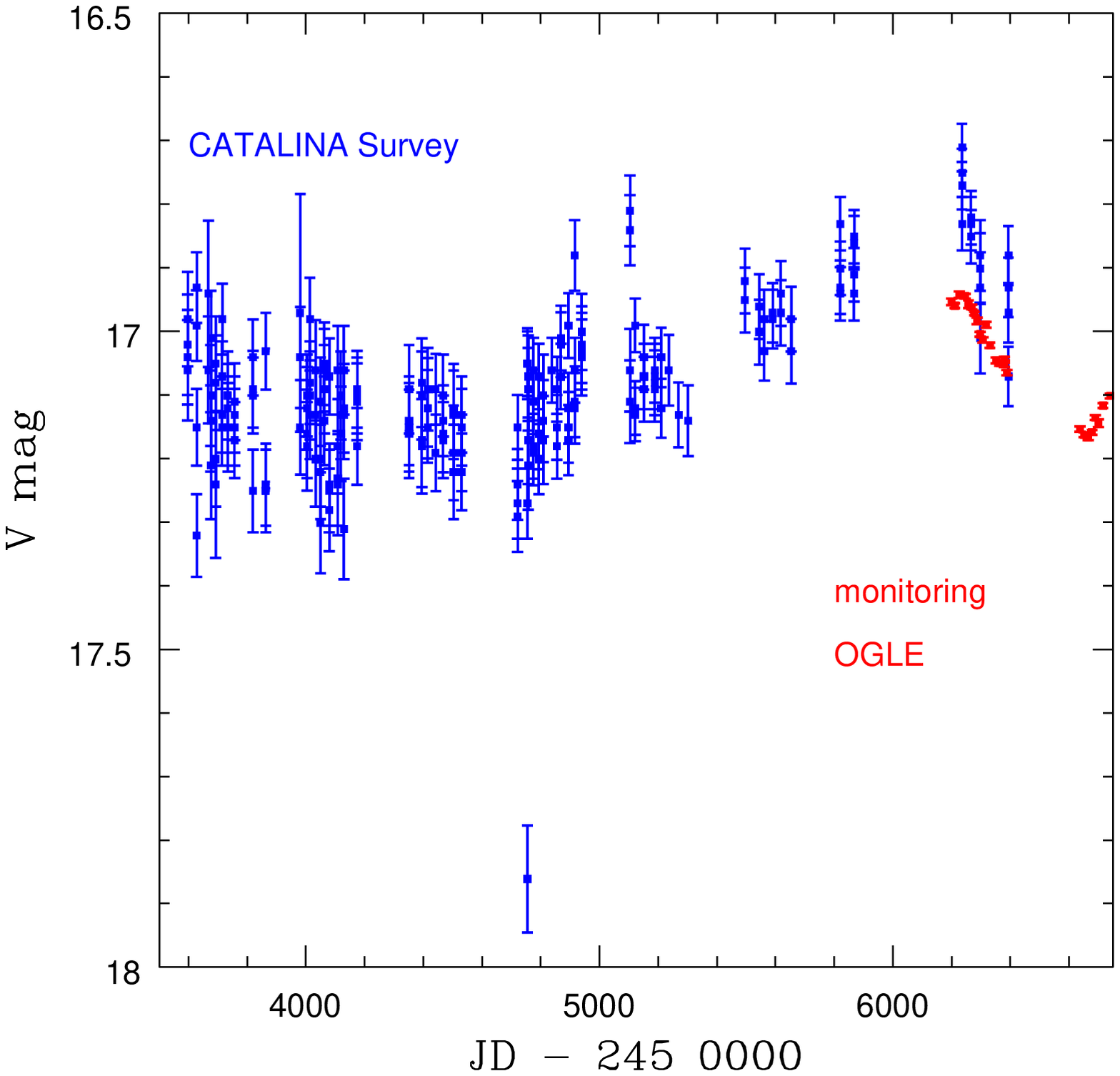}
   \caption{The V-band light curve of CTS C30.10 (middle panel) and one of the comparison star (upper panel) from OGLE monitoring, with times of spectroscopic measurements shown with vertical lines; lower panel shows the Catalina lightcurve together with the OGLE lightcurve.}
              \label{fig:lightcurve}%
    \end{figure}

The spectroscopic observations of  CTS~C30.10 were supplemented with the photometric data. 
Photometry was collected as a sub-project of the OGLE-IV
survey with the 1.3-m Warsaw Telescope and 32 CCD mosaic camera, located
at the Las Campanas Observatory, Chile. CTS~C30.10 was monitored
approximately every two weeks between September 2012 and March 2014 in
the V-band with the exposure time set to 240 seconds. Collected images
were reduced using the standard OGLE photometry pipeline (Udalski 2003).
Accuracy of single measurement of CTS~C30.10 was better than 0.01 mag.
OGLE observations reveal long term photometric changes of CTS~C30.10 in
the time scale of an order of a year. Stability of the zero point of the
photometry was verified by checking several nearby constant stars. OGLE
light curve of CTS~C30.10 is presented in Fig.~\ref{fig:lightcurve} (two upper panels). 

We supplemented these short data set with much longer but less accurate photometry 
from Catalina survey (Catalina RTS; Drake et al. 2009) which covers the last 8 years. 
The observed trends are shown in the lower panel of Fig.~\ref{fig:lightcurve}. Overall, the CATALINA lightcurve is consistent with high quality OGLE data. There is one strange outlier, well below the other data points. If real, it would imply a flux change by a factor 2 in 16 seconds which we consider unlikely. The CATALINA position of the source in this particular measurement is strangely shifted by 0.5 arc sec so probably the source was at the edge of the field of view of the instrument.

\subsection{Broad-band spectrum}

For the purpose of insight into the broad band spectral shape we also collected the data through the ASDC Sky Explorer/SED Builder web interface \footnote{http://tools.asdc.asi.it/} and GALEX View
\footnote{http://galex.stsci.edu/GalexView/}. Photometric points originate from the following data products:
WISE (Wright et al. 2010), 2MASS (Skrutskie et al. 2006),  USNO A2.0, USNO B1, GALEX (Martin et al.
2005) and
RASS (Voges et al. 1999). The data points were dereddened assuming $E(B-V) = 0.009$ from NED and applying the extinction curve of Cardelli et al. (1989) for a standard value of $R_V = 3.1$. Those points are multi-epoch data.  

\begin{figure}
   \centering

   \includegraphics[width=0.95\hsize]{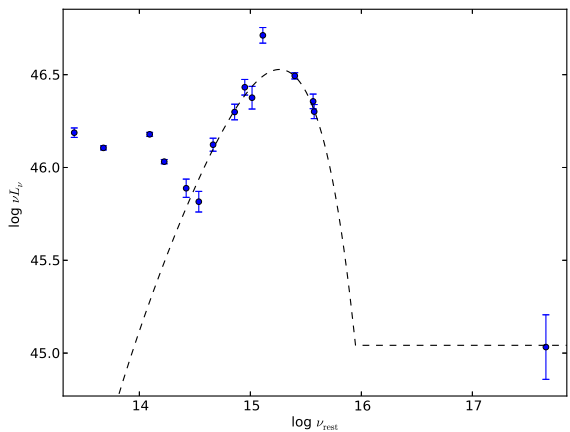}
   \caption{The time-averaged broad band spectrum of CTS C30.10, together with the parametric fit based on Eq.~\ref{eq:broad_band_fit}; The IR contribution comes from the circumnuclear dust and the host galaxy. }
              \label{fig:broad_band}%
    \end{figure}

The broad band spectrum is shown in Fig.~\ref{fig:broad_band}. This spectrum is based on observations derived in the time span of several years so it is not surprising that some points depart from the overall shape. If more data points were available at a given wavelength, like in the V band, we averaged them to show more  clearly CTS C30.10 spectral energy density (SED). The optical/UV/X-ray part of the spectrum is well represented by a simple approximation to a disk plus X-ray power law shape in the form of 
\begin{equation}
\label{eq:broad_band_fit}
F_{\nu} = A \nu^{\alpha_{uv}} \exp(-h\nu/kT_{BB})) \exp(-kT_{IR}/h\nu)  + B(\alpha_{ox})  \nu^{\alpha_x}
\end{equation}
with the following values of the parameters: $T_{BB} = 4.8\times 10^4 K$,
$\alpha_{uv} = 0.95$,
$\alpha_x = -1$,
$\alpha_{ox} = -1.55$. The choice of $\alpha_x = -1$ was arbitrary but consistent with the broad band composite shape (Laor et al. 1997a; Elvis et al. 2012). This fit, with appropriate proportionality constants A and B, is shown in Fig.~\ref{fig:broad_band} with a dashed line. It represents the contribution of an accretion disk and X-ray compact source to the total spectrum. The value of $T_{BB}$ is moderate and hints for a moderate Eddington ratio in the quasar.
The point based on the USNO catalog (B band) is much above the expected trend. This catalog is based on very old Palomar Observatory Sky Survey plates. Therefore, two effects could be responsible for the shift: either the measurement accuracy is lower than the provided errorbar, or quasar variability over the years could account for that. We do not have the precise date of that observation but the measurement is much older than most of the other SED points.


%
%

%
%


\section{Model}

We aim to model the spectrum in the relatively narrow band, between 2700 \AA~ and 2900 \AA~ rest frame, so we assume that the
spectrum consists of three components:  power-law continuum, Fe II pseudo-continuum, and Mg II line. The line has apparently much 
more complex shape than the single-Lorentzian shape found for LBQS 2113--4538 so we allow for two separate kinematic components in
the line itself, represented either as Lorentzians or as Gaussians. As an alternative, we also consider a single component in emission and 
a second component in absorption. Each of the kinematic components is modeled either as a single 
line emitted at  2800 \AA~ or as a doublet (2796.35, 2803.53; Morton 1991). The doublet ratio is usually fixed at 1:1 but we also 
check the sensitivity of the solution to this assumption and search for the best-fit ratio. 
We also vary the redshift of the object since the previously collected data used for redshift determination was not of high quality 
(Maza et al. 1993). 

The Fe II UV pseudo-continuum is modeled by a choice of theoretical and observational templates. We test classical observational template of 
the Vestergaard \& Wilkes (2001) and Tsuzuki et al. (2006) based on I Zw 1. Theoretical templates were taken from Bruhweiler and Verner (2008), each of the templates calculated for different value of the density, 
turbulent velocity, and ionization parameter $\Phi$.  Finally, we used the observational template derived by  Hryniewicz et al. (2014) as a by-product of the analysis of the LBQS 2113-4538. We allow for Fe II broadening as well as for the shift with respect to
 Mg II since Fe II was suggested to come from infalling material (Ferland et al. 2009).

\section{Results}

We analyze  five spectra of the quasar CTS C30.10 obtained with the SALT telescope between December 2012 and March 2014. Spectra are analyzed separately since the photometric lightcurve shows systematic decrease of the source luminosity between December 2012 and January 2014, later replaced by a brightening trend so we search for the corresponding variability in the spectral shape. Using the photometry, we also attempt to calibrate the spectra properly and to obtain the rms spectrum for our source. We model in detail the relatively narrow spectral range, 2700 - 2900 \AA ~ in the rest frame of the object since broader band data from SALT cannot be calibrated in a reliable way. Narrow band spectral fitting, however, is enough to have an insight into the structure of the Mg II line and its time variability.

\subsection{Separate spectra}

All spectra have high signal-to-noise ratio, in the range of 30 to 50, with the 579 pixels covering the 2700 - 2900 \AA~ rest frame band.  
All five spectra are clearly asymetric in the Mg II region. This is frequently seen in the quasar Mg II line shape spectra (e.g. Vestergaard \& Wilkes 2001, Dietrich et al. 2002, Woo 2008), and may be interpreted as multi-component nature of emission.
In particular, two-component fits to Mg II line were used by Dietrich et al. (2003), and the distinction between the Broad Line Region and the Very Broad Line Region is
done at the basis of such analysis. Shen et al. (2011) in their analysis of the quasar properties from SDSS DR7 used a single Gaussian to represent the contribution from 
the Narrow Line Region, and one to three Gaussians to describe the broad line, while Hu et al. (2008) and Wang et al. (2009) used five-parameter Gauss–Hermite series for the broad component.
So we first follow this general interpretation and study pure emission model of the chosen spectral region. However, we fit the continuum, Fe II pseudo-continuum,  and the Mg II 
line at the same time. 

The contribution from the Fe II pseudo-continuum may in principle be responsible for the apparent Mg II asymmetry.  This was the case for the previously studied quasar LBQS 2112-4538 (Hryniewicz et al. 2014), which belongs to type A sources. To check whether CTS C30.10 can be also modeled with only a single component we used several Fe II templates but for all of them the Mg II was much better represented when two kinematic components for the line were used, and the difference in $\chi^2$ between single kinematic component and two kinematic components was very large.  We give just one example of a single component fit for each observation in Table~\ref{tab:wszystkie} to illustrate this effect. The doublet nature of the two components is not directly visible in the spectra so we used the solutions with a doublet ratio 1:1 as a default. Any change of that ratio does not improve a single component fit considerably.

We thus consider two-component fits for Mg II, with both components in emission or one in emission, and one in absorption. In both cases, again we use a 1:1 doublet description of each of the kinematic components as a default.

\subsubsection{Two Mg II components in emission}
\label{sect:two_emiss}

\begin{longtab}
\begin{longtable}{lrrrrrrrrrr}
\caption{\label{tab:wszystkie}Parameters of the fits for the five individual spectra obtained with SALT between Decmber 2012 and March 2014.}  \\
\hline\hline
Model       & shape\tablefootmark{a}    &   Fe II  & slope & Mg II  & Mg II       & Mg II       & Mg II       & Mg II       & Fe II     &$\chi^2$ \\
            &  & template &       & EW     & EW          & FWHM        & EW          & FWHM        & EW        &         \\
            &    &          &       &        & comp. 1 & comp. 1 & comp. 2 & comp. 2 &           &         \\
            &     &          &       & \AA    & \AA 	 & km s$^{-1}$ &  \AA        & km s$^{-1}$ & \AA       &         \\
\hline                        
\endfirsthead
\caption{continued.}\\
\hline\hline      
Model       & shape\tablefootmark{a}    &   Fe II  & slope & Mg II  & Mg II       & Mg II       & Mg II       & Mg II       & Fe II     &$\chi^2$ \\
            &  & template &       & EW     & EW          & FWHM        & EW          & FWHM        & EW        &         \\
            &    &          &       &        & comp. 1 & comp. 1 & comp. 2 & comp. 2 &           &         \\
            &     &          &       & \AA    & \AA 	 & km s$^{-1}$ &  \AA        & km s$^{-1}$ & \AA       &         \\
\hline                        
\endhead
\hline
\endfoot

Obs. 1 \\
\hline

A           &    DL        & temp 01   &-1.4  & 35.08   & 14.4        & 3243        & 20.7        & 7479     &  8.35      &  300.47  \\ 
B           &    DL        & temp 02   &-1.6  & 29.52   & 12.9        & 2812        & 16.6        & 6870     &  6.36      &  273.62  \\ 
C           &    DL        & temp 03   &-1.3  & 25.00   & 16.8        & 2775        &  8.2        & 3145     &  8.95      &  233.64  \\
D           &    DL        & temp 04   &-1.3  & 25.27   & 17.0        & 2825        &  8.3        & 3239     &  8.69      &  232.89  \\
E           &    DL        & temp 05   &-1.3  & 24.89   & 17.0        & 2846        &  7.9        & 3142     &  7.84      &  236.32  \\
F           &    DL        & temp 06   &-1.2  & 25.33   & 17.1        & 2912        &  8.2        & 3351     &  7.86      &  234.16  \\
G           &    DL        & temp 07   &-1.2  & 24.90   & 16.9        & 2918        &  8.0        & 3408     &  5.56      &  237.73  \\
H           &    DL        & temp 08   &-1.2  & 25.09   & 17.0        & 2934        &  8.1        & 3420     &  6.36      &  237.20  \\
I           &    DL        & temp 09   &-1.2  & 25.23   & 17.0        & 2850        &  8.2        & 3251     &  8.79      &  232.58  \\
J           &    DL        & temp 10   &-1.2  & 25.56   & 17.3        & 2912        &  8.3        & 3345     &  8.80      &  232.45  \\ 
K           &    DL        & temp 11   &-1.1  & 25.97   & 17.6        & 2931        &  8.4        & 3298     & 10.54      &  229.89  \\
L           &    DL        & temp 12   &-1.2  & 25.47   & 17.1        & 2909        &  8.3        & 3433     &  7.27      &  233.22  \\ 
M           &    DL        & temp 13   &-1.2  & 25.63   & 17.1        & 2871        &  8.5        & 3470     &  6.98      &  230.78  \\
N           &    SL        & temp 13   &-0.7  & 26.37   &  -          & 5022        &    -        &   -      &  2.31E-04  & 1155.71  \\
O           &    DG        & temp 13   &-0.9  & 37.89   & 25.50       & 17344       & 12.4        & 4259     &  0.35      &  687.22  \\
P           &    SG        & temp 13   &-0.6  & 18.37   &  -          & 5515        &    -        &   -      &  4.92E-02  & 1593.88  \\
Q           &    DL        & temp 14   &-1.2  & 25.52   & 16.9        & 2878        &  8.7        & 3520     &  7.40      &  230.13  \\
R           &    DL        & temp 15   &-1.2  & 25.30   & 17.4        & 2950        &  8.0        & 3251     &  8.26      &  234.82  \\
S           &    DL        & temp 16   &-1.4  & 24.83   & 16.8        & 2962        &  8.1        & 3604     &  3.79      &  281.96  \\

\hline                                   
\hline                        
Obs. 2 \\
\hline
A           &    DL        & temp 01   &-1.1  & 28.96   & 23.4       & 3725     &  5.6       & 2991     &  6.11      &  447.57  \\ 
B           &    DL        & temp 02   &-1.1  & 25.73   & 17.5       & 2990     &  8.2       & 3606     &  9.34      &  430.38  \\ 
C           &    DL        & temp 03   &-1.2  & 25.75   & 16.6       & 2771     &  9.1       & 3144     &  9.77      &  426.84  \\
D           &    DL        & temp 04   &-1.2  & 25.99   & 16.8       & 2825     &  9.2       & 3225     &  9.45      &  435.04  \\
E           &    DL        & temp 05   &-1.2  & 25.61   & 16.9       & 2868     &  8.7       & 3172     &  8.16      &  440.35  \\
F           &    DL        & temp 06   &-1.1  & 26.09   & 17.1       & 2931     &  9.0       & 3334     &  8.59      &  433.52  \\
G           &    DL        & temp 07   &-1.1  & 25.44   & 16.8       & 2940     &  8.7       & 3369     &  5.72      &  431.87  \\
H           &    DL        & temp 08   &-1.1  & 25.76   & 17.0       & 2962     &  8.9       & 3409     &  6.72      &  438.81  \\
I           &    DL        & temp 09   &-1.1  & 26.04   & 17.0       & 2828     &  9.2       & 3181     & 10.53      &  421.85  \\
J           &    DL        & temp 10   &-1.1  & 26.39   & 17.2       & 2925     &  9.2       & 3316     &  9.93      &  431.14  \\ 
K           &    DL        & temp 11   &-1.0  & 26.93   & 17.6       & 2937     &  9.3       & 3238     & 12.36      &  424.05  \\
L           &    DL        & temp 12   &-1.1  & 26.24   & 17.0       & 2928     &  9.2       & 3422     &  7.92      &  434.13  \\ 
M           &    DL        & temp 13   &-1.1  & 26.51   & 16.9       & 2850     &  9.6       & 3438     &  8.38      &  422.49  \\
N           &    SL        & temp 13   &-0.5  & 28.18   &   -        & 5316     &    -       &   -      &  2.30E-04  & 1017.32  \\
O           &    DG        & temp 13   &-0.8  & 20.85   & 14.1       & 3326     &  6.7       & 3274     & 12.78      &  469.17  \\
P           &    SG        & temp 13   &-0.5  & 19.31   &   -        & 5703     &    -       &   -      &  2.82E-04  & 1247.38  \\
Q           &    DL        & temp 14   &-1.1  & 25.85   & 16.9       & 2956     &  9.0       & 3450     &  6.38      &  430.29  \\
R           &    DL        & temp 15   &-1.1  & 26.13   & 17.4       & 2978     &  8.7       & 3238     &  9.12      &  435.09  \\
S           &    DL        & temp 16   &-1.3  & 25.42   & 16.2       & 2865     &  9.2       & 3461     &  5.87      &  446.53  \\
\hline                                   
\hline                        
Obs. 3 \\
\hline
A           &    DL        & temp 01   &-1.2  & 29.47  & 25.4       & 3565     &  4.1       & 2098     &  5.65      &  234.74  \\ 
B           &    DL        & temp 02   &-1.1  & 26.78  & 21.0       & 2862     &  5.8       & 2498     & 13.63      &  208.81  \\ 
C           &    DL        & temp 03   &-1.3  & 26.54  & 19.4       & 2718     &  7.2       & 2401     & 10.36      &  202.99  \\
D           &    DL        & temp 04   &-1.3  & 26.97  & 19.6       & 2725     &  7.4       & 2433     & 11.47      &  204.55  \\
E           &    DL        & temp 05   &-1.3  & 26.70  & 19.6       & 2721     &  7.2       & 2339     & 11.36      &  207.27  \\
F           &    DL        & temp 06   &-1.1  & 27.06  & 19.8       & 2811     &  7.3       & 2484     & 11.14      &  204.57  \\
G           &    DL        & temp 07   &-1.1  & 26.19  & 19.4       & 2846     &  6.8       & 2529     &  6.92      &  207.22  \\
H           &    DL        & temp 08   &-1.1  & 26.51  & 19.6       & 2865     &  6.9       & 2536     &  8.13      &  207.98  \\
I           &    DL        & temp 09   &-1.1  & 26.87  & 19.6       & 2750     &  7.3       & 2423     & 11.80      &  200.30  \\
J           &    DL        & temp 10   &-1.1  & 27.57  & 20.0       & 2790     &  7.6       & 2470     & 13.49      &  203.87  \\ 
K           &    DL        & temp 11   &-1.0  & 28.29  & 20.6       & 2812     &  7.7       & 2417     & 16.40      &  200.53  \\
L           &    DL        & temp 12   &-1.1  & 26.83  & 19.8       & 2859     &  7.1       & 2561     &  8.47      &  206.03  \\ 
M           &    DL        & temp 13   &-1.1  & 27.45  & 19.7       & 2750     &  7.8       & 2636     & 10.21      &  202.69  \\
N           &    SL        & temp 13   &-0.7  & 28.91  &   -        & 4900     &    -       &   -      &  2.64E-04  &  494.53  \\
O           &    DG        & temp 13   &-1.0  & 47.39  & 34.2       & 18130    & 13.2       & 4039     &  0.88      &  371.37  \\
P           &    SG        & temp 13   &-0.6  & 20.35  &   -        & 5412     &    -       &   -      &  0.79      &  654.71 \\
Q           &    DL        & temp 14   &-1.2  & 26.86  & 19.5       & 2825     &  7.3       & 2636     &  8.77      &  204.81  \\
R           &    DL        & temp 15   &-1.1  & 27.26  & 20.1       & 2834     &  7.1       & 2383     & 12.69      &  205.16  \\
S           &    DL        & temp 16   &-1.3  & 26.12  & 18.9       & 2756     &  7.2       & 2595     &  7.25      &  223.20  \\
\hline                        

Obs. 4 \\
\hline
A           &    DL        & temp 01   &-1.8  & 41.92  & 10.6       & 2815     & 31.4       & 7509     & 10.02      &  683.98  \\ 
B           &    DL        & temp 02   &-1.5  & 29.38  & 19.7       & 2806     &  9.7       & 3578     & 25.02      &  601.94  \\ 
C           &    DL        & temp 03   &-1.8  & 28.47  & 17.8       & 2698     & 10.7       & 3209     & 14.63      &  609.77  \\
D           &    DL        & temp 04   &-1.8  & 28.77  & 18.1       & 2791     & 10.7       & 3350     & 13.49      &  611.58  \\
E           &    DL        & temp 05   &-1.8  & 28.44  & 18.1       & 2787     & 10.3       & 3206     & 13.62      &  620.40  \\
F           &    DL        & temp 06   &-1.6  & 29.03  & 18.4       & 2921     & 10.7       & 3525     & 12.61      &  612.65  \\
G           &    DL        & temp 07   &-1.6  & 28.32  & 17.9       & 2903     & 10.4       & 3594     &  9.43      &  617.59  \\
H           &    DL        & temp 08   &-1.6  & 28.65  & 18.2       & 2953     & 10.5       & 3619     & 10.37      &  620.76  \\
I           &    DL        & temp 09   &-1.6  & 28.83  & 18.2       & 2831     & 10.6       & 3378     & 14.01      &  605.40  \\
J           &    DL        & temp 10   &-1.5  & 29.60  & 18.5       & 2887     & 11.1       & 3488     & 15.32      &  609.19  \\ 
K           &    DL        & temp 11   &-1.5  & 30.36  & 19.0       & 2921     & 11.4       & 3413     & 18.35      &  605.11  \\
L           &    DL        & temp 12   &-1.6  & 29.21  & 18.3       & 2915     & 10.9       & 3631     & 11.58      &  612.29  \\ 
M           &    DL        & temp 13   &-1.6  & 29.47  & 18.3       & 2850     & 11.2       & 3666     & 11.30      &  604.04  \\
N           &    SL        & temp 13   &-1.0  & 31.44  &   -        & 5644     &   -        &   -      &  2.24E-04  & 1738.84  \\
O           &    DG        & temp 13   &-1.3  & 23.38  & 15.4       & 3288     &  7.9       & 3389     & 17.98      &  682.53  \\
P           &    SG        & temp 13   &-0.9  & 22.05  &   -        & 6213     &   -        &   -      &  0.00      & 2205.63  \\
Q           &    DL        & temp 14   &-1.6  & 29.33  & 17.8       & 2846     & 11.5       & 3716     & 12.34      &  605.93  \\
R           &    DL        & temp 15   &-1.5  & 29.17  & 18.8       & 2943     & 10.4       & 3347     & 14.62      &  615.34  \\
S           &    DL        & temp 16   &-1.8  & 28.27  & 17.2       & 2881     & 11.1       & 3839     &  7.91      &  711.19  \\
\hline                                   
\hline                        
Obs. 5 \\
\hline
A           &    DL        & temp 01   &-1.9  & 42.63  &  8.1       & 2193     & 34.5       & 6988     & 10.30      &  307.59  \\ 
B           &    DL        & temp 02   &-1.6  & 29.53  & 20.6       & 2750     &  8.9       & 3363     & 20.66      &  241.24  \\ 
C           &    DL        & temp 03   &-1.9  & 28.97  & 19.0       & 2639     & 10.0       & 3022     & 13.38      &  243.12  \\
D           &    DL        & temp 04   &-1.9  & 29.25  & 19.2       & 2698     & 10.1       & 3116     & 12.82      &  241.99  \\
E           &    DL        & temp 05   &-1.8  & 28.81  & 19.2       & 2709     &  9.6       & 2997     & 12.25      &  248.03  \\
F           &    DL        & temp 06   &-1.6  & 29.33  & 19.4       & 2796     &  9.9       & 3197     & 12.13      &  235.25  \\
G           &    DL        & temp 07   &-1.7  & 28.66  & 18.9       & 2765     &  9.7       & 3241     &  9.45      &  235.95  \\
H           &    DL        & temp 08   &-1.7  & 28.89  & 19.2       & 2821     &  9.7       & 3253     &  9.96      &  237.44  \\
I           &    DL        & temp 09   &-1.6  & 29.25  & 19.2       & 2705     & 10.0       & 3084     & 13.94      &  234.95  \\
J           &    DL        & temp 10   &-1.6  & 29.72  & 19.6       & 2781     & 10.2       & 3181     & 13.96      &  235.08  \\ 
K           &    DL        & temp 11   &-1.6  & 30.47  & 20.2       & 2812     & 10.3       & 3142     & 16.67      &  235.54  \\
L           &    DL        & temp 12   &-1.7  & 29.52  & 19.4       & 2778     & 10.1       & 3261     & 11.63      &  235.43  \\ 
M           &    DL        & temp 13   &-1.7  & 29.82  & 19.3       & 2728     & 10.5       & 3338     & 11.21      &  232.07  \\
N           &    SL        & temp 13   &-1.1  & 30.69  &   -        & 5122     &   -        &   -      &  2.85E-04  & 1134.71  \\
O           &    DG        & temp 13   &-1.4  & 23.93  & 16.18      & 3142     &  7.8       & 3106     & 18.22      &  268.18  \\
P           &    SG        & temp 13   &-1.0  & 21.51  &   -        & 5713     &   -        &   -      &  2.41E-09  & 1555.81  \\
Q           &    DL        & temp 14   &-1.7  & 29.66  & 19.0       & 2715     & 10.8       & 3366     & 12.41      &  231.64  \\
R           &    DL        & temp 15   &-1.6  & 29.41  & 19.8       & 2831     &  9.6       & 3075     & 13.35      &  238.80  \\
S           &    DL        & temp 16   &-1.9  & 28.29  & 18.5       & 2800     &  9.8       & 3426     &  6.64      &  313.02  \\
\hline                        

\end{longtable}
\tablefoot{Templates are numbered in the followng way: temp 1 is the observational template from Vestergaard \& Wilkes (2001), temp 2 is the observational template of Tsuzuki et al. (2006), templates 3 - 15 are theoretical templates of Bruhweiler \& Verner (2008), desinated as d11-m20-20.5-735, d11-m30-20-5-735, d11-m20-21-735, d10-5-m20-20-5, d11-m05-20-5, d11-m10-20-5, d11-m20-20-5, d11m30-20-5, d11-m50-20-5, d11-5-m20-20-5, d12-m20-20-5, d11-m20-20,d11-m20-21 in their paper, and template 16 is the observational template from Hryniewicz et al. (2014). }\\
\tablefoottext{a}{Line shape: single/double gaussian or lorentzian.}
\end{longtab}

   \begin{figure}
   \centering
\vskip - 3.0 true cm
   \includegraphics[width=0.9\hsize]{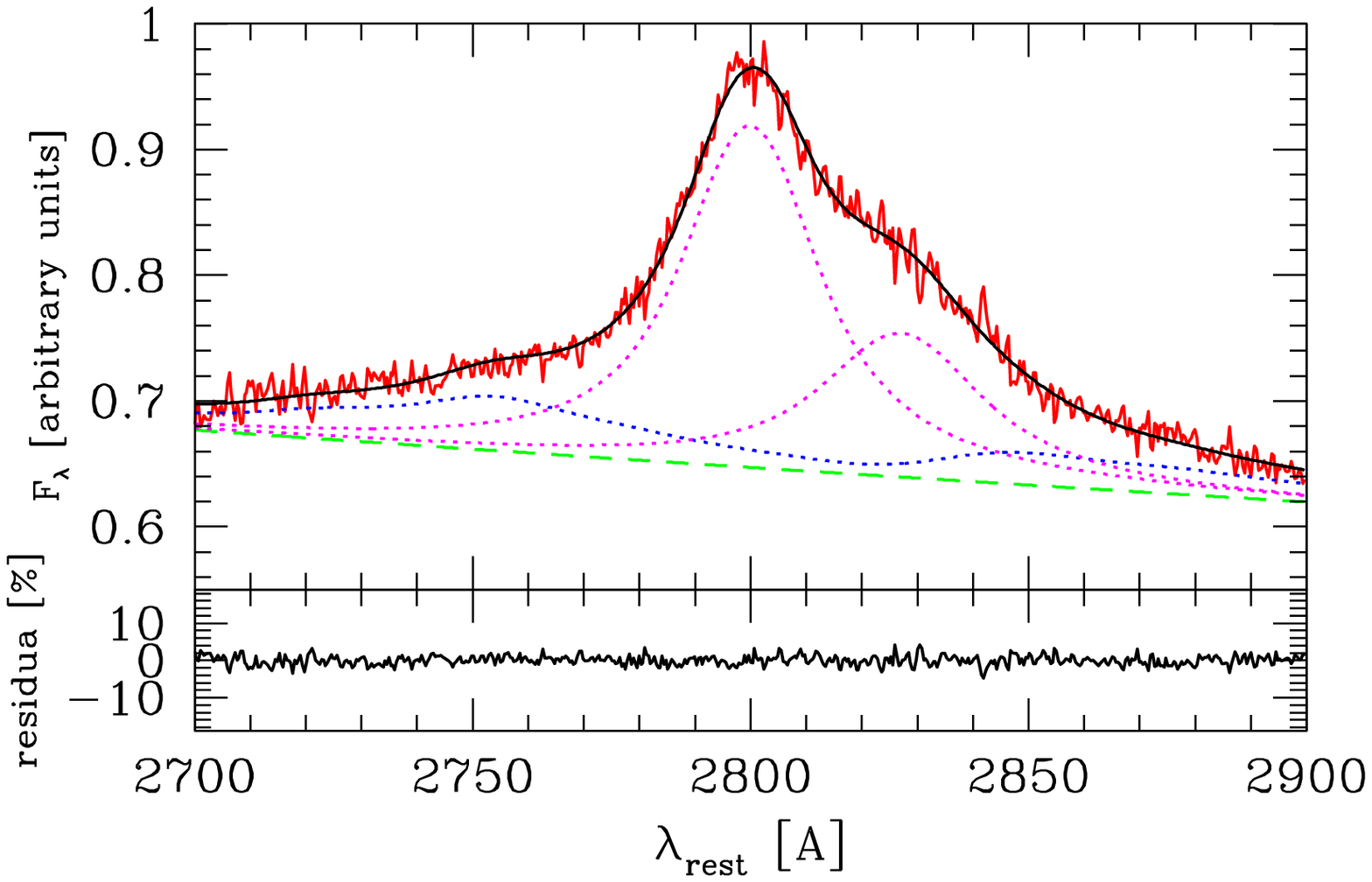}
\vskip - 4.0 true cm
  \includegraphics[width=0.9\hsize]{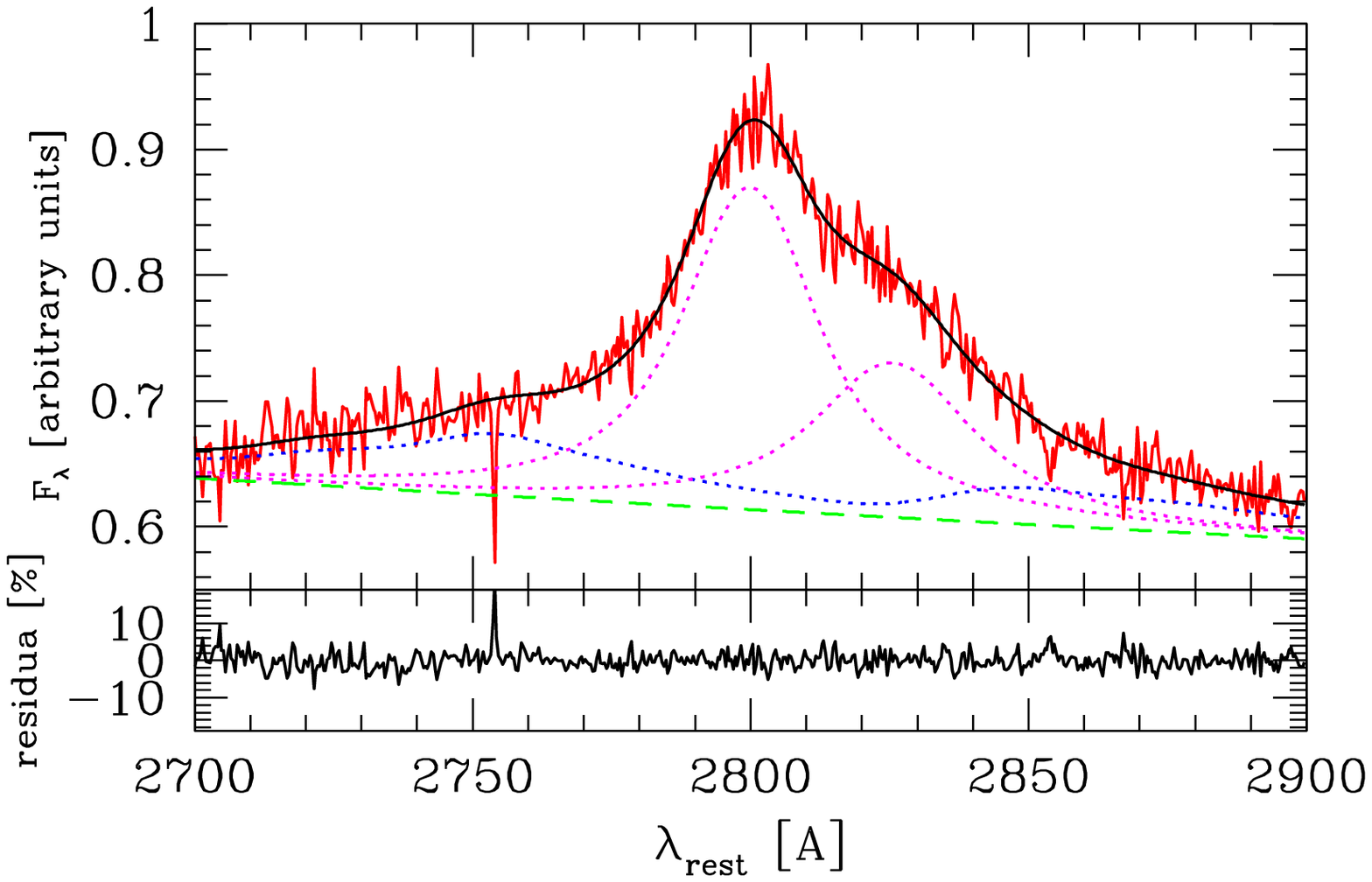}
\vskip - 4.0 true cm
  \includegraphics[width=0.9\hsize]{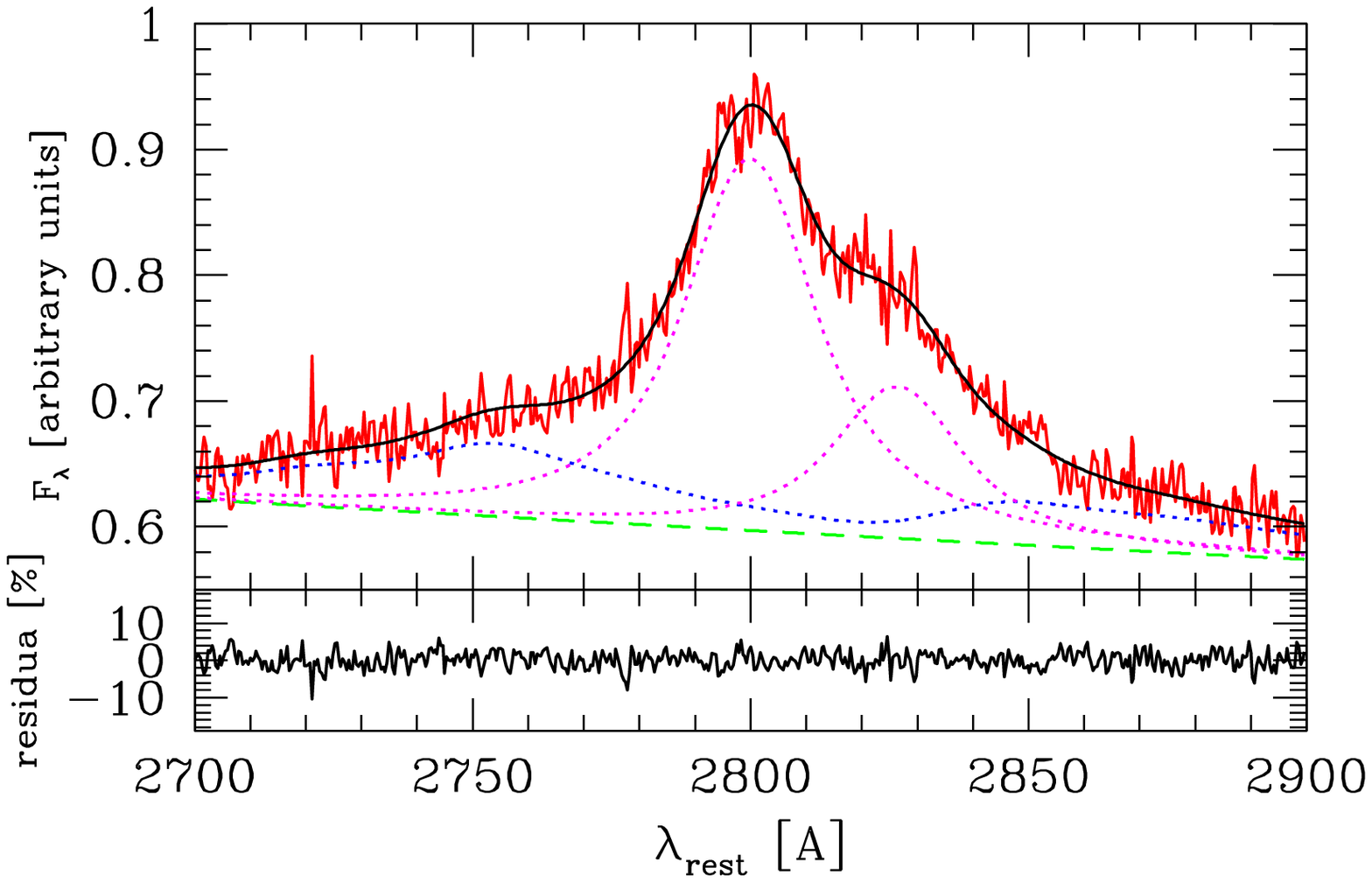}
\vskip - 4.0 true cm
   \includegraphics[width=0.9\hsize]{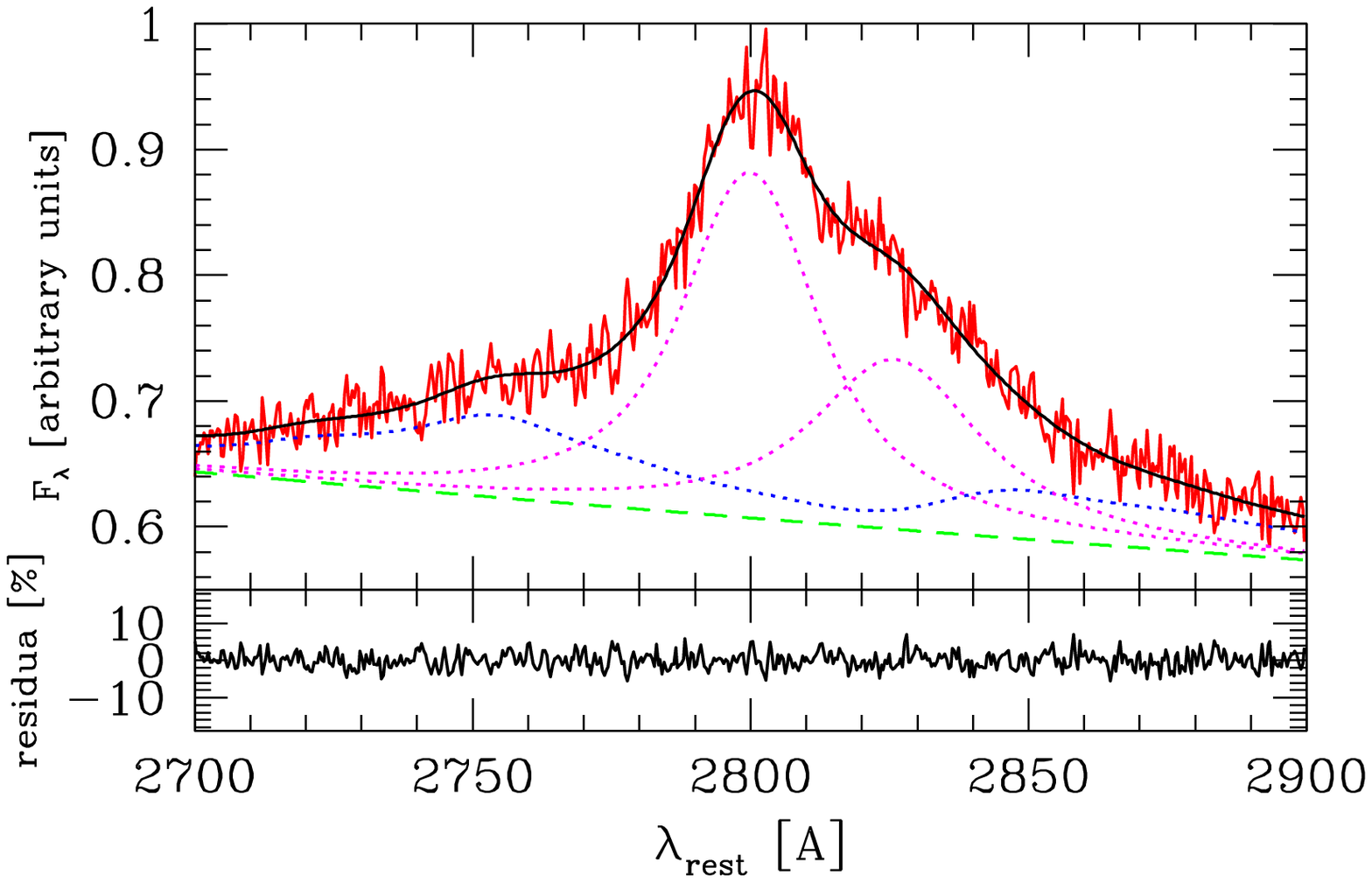}
\vskip - 4.0 true cm
  \includegraphics[width=0.9\hsize]{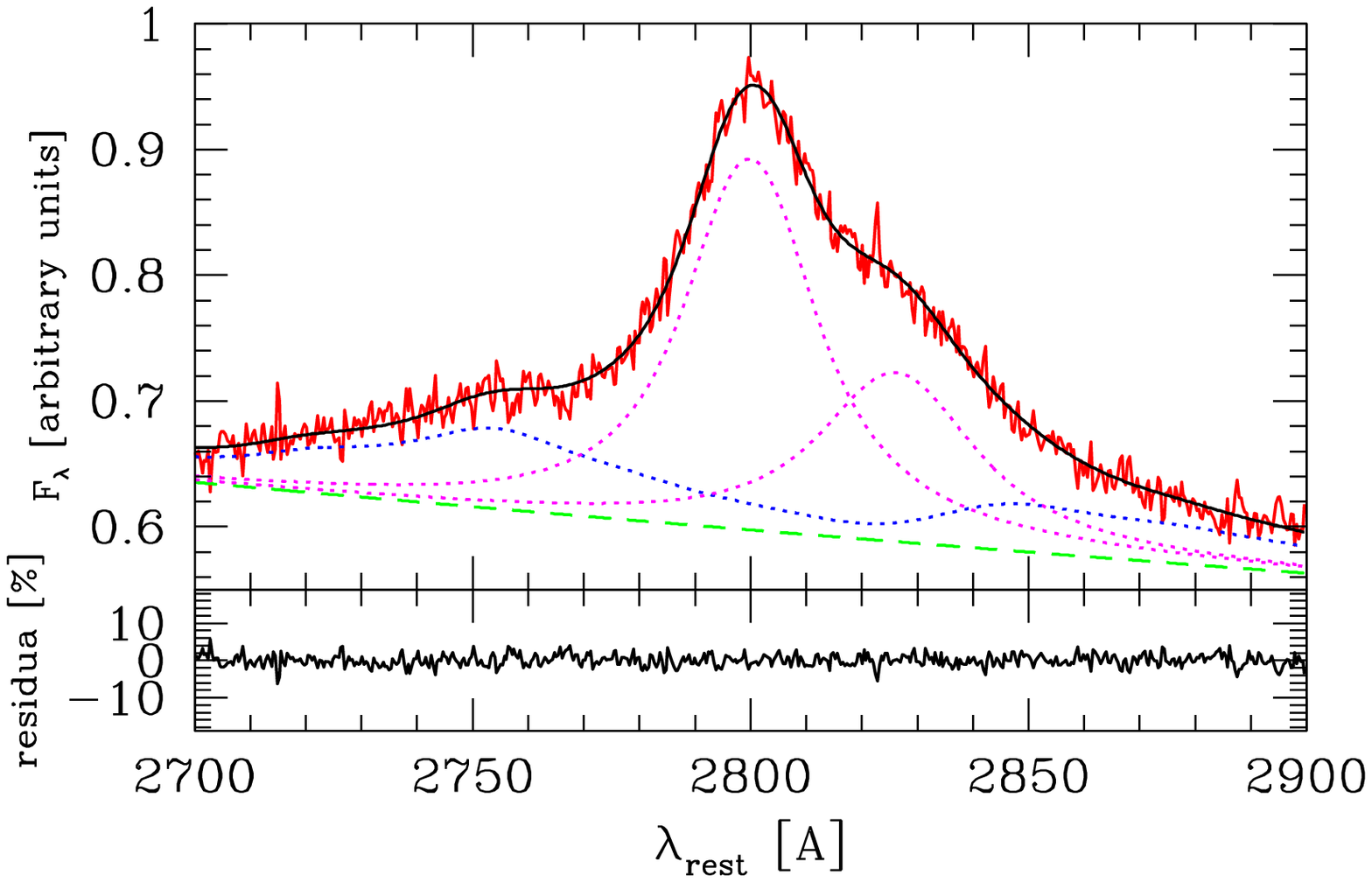}
\vskip - 0.5 true cm
   \caption{The best fit and residuals for five observations, model M (see Table~\ref{tab:wszystkie} for model parameters); continuous lines show the model and the data, dashed lines give the underlying power law, the dotted blue line represents the kinematically blurred Fe II pseudo-continuum, and dotted magenta lines mark the two kinematic components of the Mg II.}
              \label{fig:widma}%
    \end{figure}

\begin{table*}
\caption{Summary table of the fits of Mg II with errors for the template 13  and two Lorentzian components in emission  for five individual spectra obtained with SALT between December 2012 and March 2014.}   
\label{tab:wybrane}      
\centering                          
\begin{tabular}{l r r r r r r r r r}        
\hline\hline      
Obs.     & Mg II  & Mg II   &Mg II       & Mg II       & Mg II       & Mg  II   &  Fe II   &   Fe II\\
         &  EW    &    L    &FWHM        & EW          &  FWHM       & EW       &   EW     &    L   \\
         &  total & total   & comp. 1 & comp. 1 & comp. 2 &  comp. 2  \\
         &  \AA   & $10^{-14}$erg s$^{-1}$ cm$^{-2}$         &  km s$^{-1}$  & \AA         & km s$^{-1}$  & \AA   &  \AA &  $10^{-14}$erg s$^{-1}$ cm$^{-2}$ \\
\hline                        
1           &  $25.63^{+0.22}_{-0.13}$ & $2.95^{0.06}_{-0.05}$ &  $2871^{+170}_{-110}$ &  $ 17.1^{+1.0}_{-0.8}$  & $3470^{+380}_{-330}$  & $8.5^{+0.9}_{-0.7}$  & $6.98^{+0.71}_{-2.07}$  & $0.80^{+0.08}_{-0.23}$\\ 
2           &  $26.51^{+0.24}_{-0.14}$ & $2.89^{0.06}_{-0.05}$  & $2850^{+210}_{-140}$ &  $ 16.9^{+1.3}_{-1.0}$  & $3438^{+510}_{-400}$  & $9.6^{+1.2}_{-0.9}$  & $8.38^{+1.03}_{-3.41}$  & $0.91^{+0.11}_{-0.37}$  \\ 
3           &  $27.45^{+0.45}_{-0.07}$ & $2.93^{0.08}_{-0.05}$  & $2750^{+350}_{-40}$ &   $ 19.7^{+1.8}_{-1.0}$  & $2636^{+600}_{-400}$  & $7.8^{+1.5}_{-1.0}$  & $10.21^{+2.05}_{-2.03}$  &  $1.08^{+0.23}_{-0.23}$ \\
4           &  $29.47^{+0.22}_{-0.09}$ & $2.94^{0.06}_{-0.05}$  & $2850^{+140}_{-140}$ &  $ 18.3^{+1.0}_{-0.8}$  & $3666^{+390}_{-320}$  & $11.2^{+1.0}_{-0.7}$ & $11.30^{+1.89}_{-0.86}$ &  $1.12^{+0.20}_{-0.09}$  \\
5           &  $29.82^{+0.33}_{-0.60}$ & $2.76^{0.06}_{-0.08}$  & $2728^{+170}_{-120}$ &  $ 19.3^{+1.0}_{-1.0}$  & $3338^{+550}_{-410}$  & $10.5^{+0.8}_{-0.8}$ & $11.21^{+0.98}_{-2.31}$ &  $1.03^{+0.09}_{-0.21}$  \\
\hline                                   

\end{tabular}
\end{table*}


We consider two kinematic components for the emission of Mg II, described either as Gaussians or Lorentzians. However, we described the underlying Fe II emission as a single kinematic component.
 Fe II component was assumed to have the same  
redshift as the first (blue) Mg II component, and the best fits in almost all cases were obtained for redshift 0.9000, with the accuracy of the redshift measurement of 0.0004. This redshift, centered at the first component, is clearly lower than $z = 0.910$ determined by Maza et al. (1993).  The  shift of the second component was a free parameter of the model.  

The best fits were obtained always for a two-Lorentzian model of the Mg II shape. Fits based on two Gaussians have $\chi^2$ much higher than two Lorentzian fits, as we illustrate in Table~\ref{tab:wszystkie} for a few cases.  We thus further constrain our discussion to the two Lorentzian case. The fits depend to some extent on the choice of the Fe II template. In each observation another template gives the best fit. However, for each observation a number of templates give acceptable results within 1 sigma error.  The values of the EW for Mg II and Fe II depend to some extent on the template, although not much for templates giving the smallest $\chi^2$ values. 

In order to study the time-dependence of the parameters it is better to select a fixed Fe II template. The template 13 (theoretical template d12-m20-20-5 of Bruhweiler \& Verner 2008), with the Gaussian broadening, for a half-width of 900 km s$^{-1}$ is always in the group of acceptable templates for
all five observations so we consider this template as the best. We thus show in Table~\ref{tab:wybrane} fits for this choice of the Fe II for all five observations, including the parameter errors.  These fits to the five spectra are shown in Fig.~\ref{fig:widma}. 

The source shows significant evolution. The total equivalent width (EW) of Mg II line increases monotonically with time as the continuum gets fainter. Both kinematic components show significant increase. The ratio between the two components, however, does not show a clear trend. It varies from 0.37 to 0.58 in various data sets, but the error of the component ratio is larger than the error of the total line EW, of order of 0.08,  and such variations are in fact consistent with a constant fraction within 2 sigma error.  The line width of the first component is constant within the measurement accuracy. The second component seems to vary more: it is significantly narrower ($\sim 2600$ instead of $\sim 3500$ km s$^{-1}$) in the observation 3  and somewhat lower in observation 5 that in the other three data sets. This behavior is not accompanied by any other considerable change in the fit parameters. However, the error of the FWHM of the second component is also larger so the change cannot be considered as significant. The relative shift between the two components did not vary:  for all five observations the shift was consistent with the mean value of 2780 km s$^{-1}$ within 1 sigma error.  Fe II equivalent width given in Tables~\ref{tab:wszystkie} and ~\ref{tab:wybrane} is measured between 2700 and 2900 \AA~ rest frame. This value is rising together with the Mg II line. 

Since the continuum decreases, as seen in Fig.~\ref{fig:lightcurve}, we checked whether this change simply reflects the change in the continuum or it reflects the intrinsic change in the Mg II and Fe II luminosity. The errors are large, but the change of Mg II is nevertheless significant (see Table~\ref{tab:wybrane}). The luminosity is consistent with being constant for the first four observations, followed by a decrease in observation 5. The decrease is not large, only by 6 \% but it is highly significant. The Fe II luminosity seems rising, followed by a decrease in observation 5 but errors are large. The change in the line intensity is four times smaller than the change in the continuum (23 \%, as seen in OGLE monitoring). Thus the variability we see in Mg II is smaller than observed by  Woo (2008) but consistent with the average Mg II efficiency of $\eta \sim 0.2$ (Goad et al. 1999; Korista \& Goad 2004; Kokubo et al. 2014). 

Results given in Table~\ref{tab:wybrane} were obtained for a fixed 1:1 ratio for the Mg II doublet. We tested this assumption for observation 5. The best fit value was obtained for the 1.6:1 ratio, and the redshift 0.90052, but all values of the doublet ratio are acceptable within 2 sigma error. The EW(Mg II) line only weakly depend on the value of that ratio and changed from 29.82 \AA~ for 1:1 ratio to  29.90 \AA~ for 1.6:1. 

Fits given in Table~\ref{tab:wszystkie} are based on the assumption that there is no net shift of the Fe II with respect to the blue component of Mg II. We tested this assumption in detail for Observation 5, model M. If the relative velocity of Fe II and Mg II is treated as another free parameter of the model, the minimum is reached for the value of 450 km s$^{-1}$ (the $\chi^2$ dropped by 6.8, and all values between 180 km s$^{-1}$ and 820 km s$^{-1}$ are acceptable within 1 sigma error. Introduction of this relative velocity between the Fe II and Mg II did not change the EW(Mg II) significantly, therefore we did not repeat all the fitting with this new parameter included since this would mean the introduction of still one more free parameter.  

Since there are two Mg II components we also checked whether two components are needed to describe Fe II contribution. Since we did not want to introduce too many additional free parameters we assumed the second Fe II kinematic component is in the same proportion to the first one as the two Mg II components, and we assumed the same velocity shift for the second (red) component of Fe II. The fit was slightly worse (the $\chi^2$ was higher by 21) so the second Fe II component was clearly ruled out. This can be simply understood looking at the shape of the spectrum: there is a clear trace of the Fe II peak at $\sim 2750$ in the observed spectrum as well as in the templates, and a composition of two shifted templates smears this feature too much in the model.  

\subsubsection{Contribution from the Narrow Line Region}

In the fits above we neglected any possible contribution from the Narrow Line Region (NLR) to the Mg II line. The spectra do not seem to show a narrow top but in principle, the NLR contribution can be still hidden in the Mg II complex profile due to the relatively low spectral resolution. Therefore, we considered fits with the NLR contribution on the top of the two-component broad Mg II line. We fixed the FWHM of the narrow line at 600 km s$^{-1}$, and modeled the line as a single Gaussian at arbitrary position. However, the $\chi^2$ always dropped. Any contribution from the NLR at the level higher than 2\% is ruled out. Therefore, CTS C30.10 in this respect is similar to the type A source LBQS 2113-4538, where also the separate NLR contribution to Mg II was not detected (Hryniewicz et al. 2014).

\subsubsection{Emission/absorption interpretation}

   \begin{figure}
   \centering
\vskip - 3.0 true cm
   \includegraphics[width=0.9\hsize]{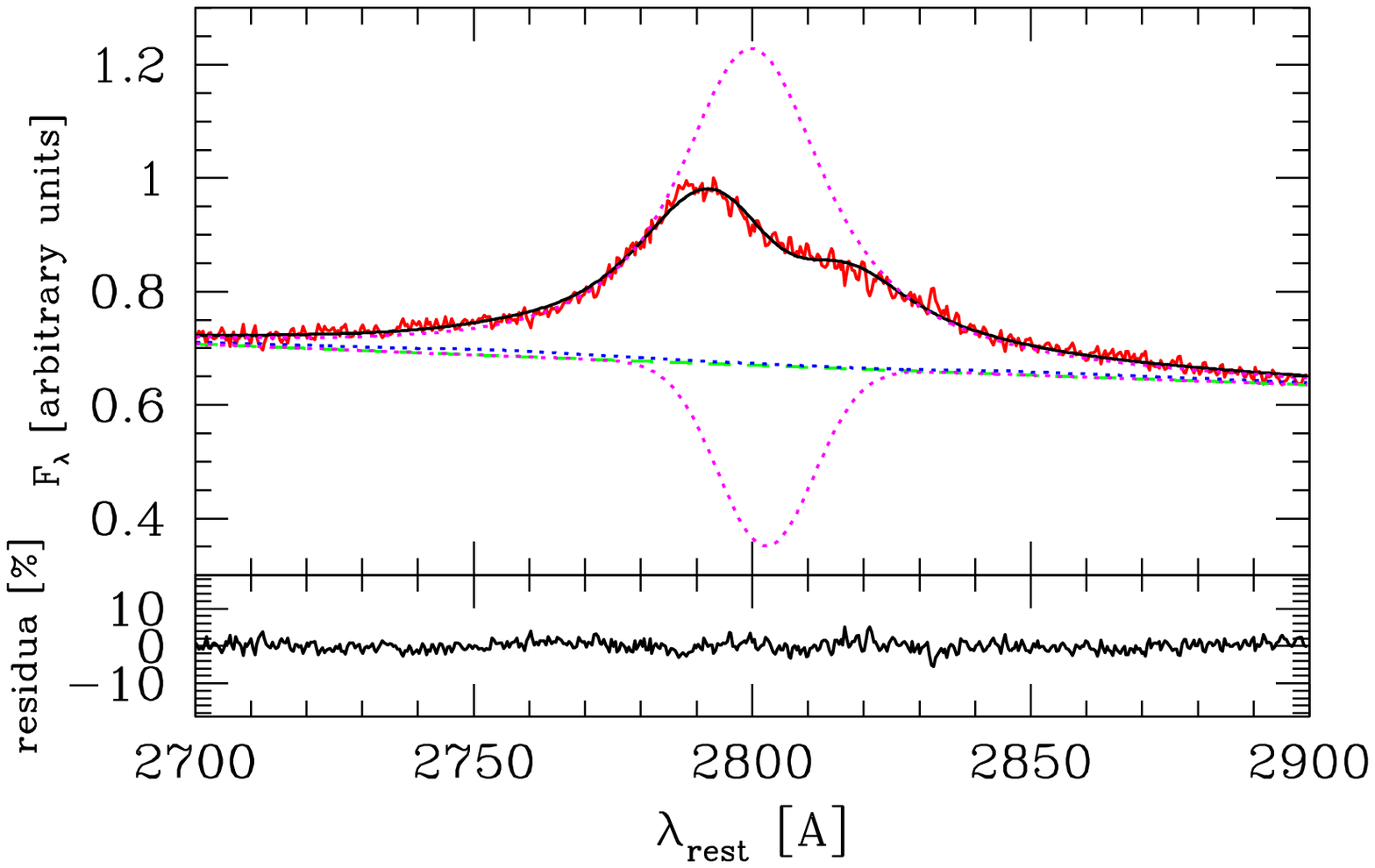}
\vskip - 4.0 true cm
  \includegraphics[width=0.9\hsize]{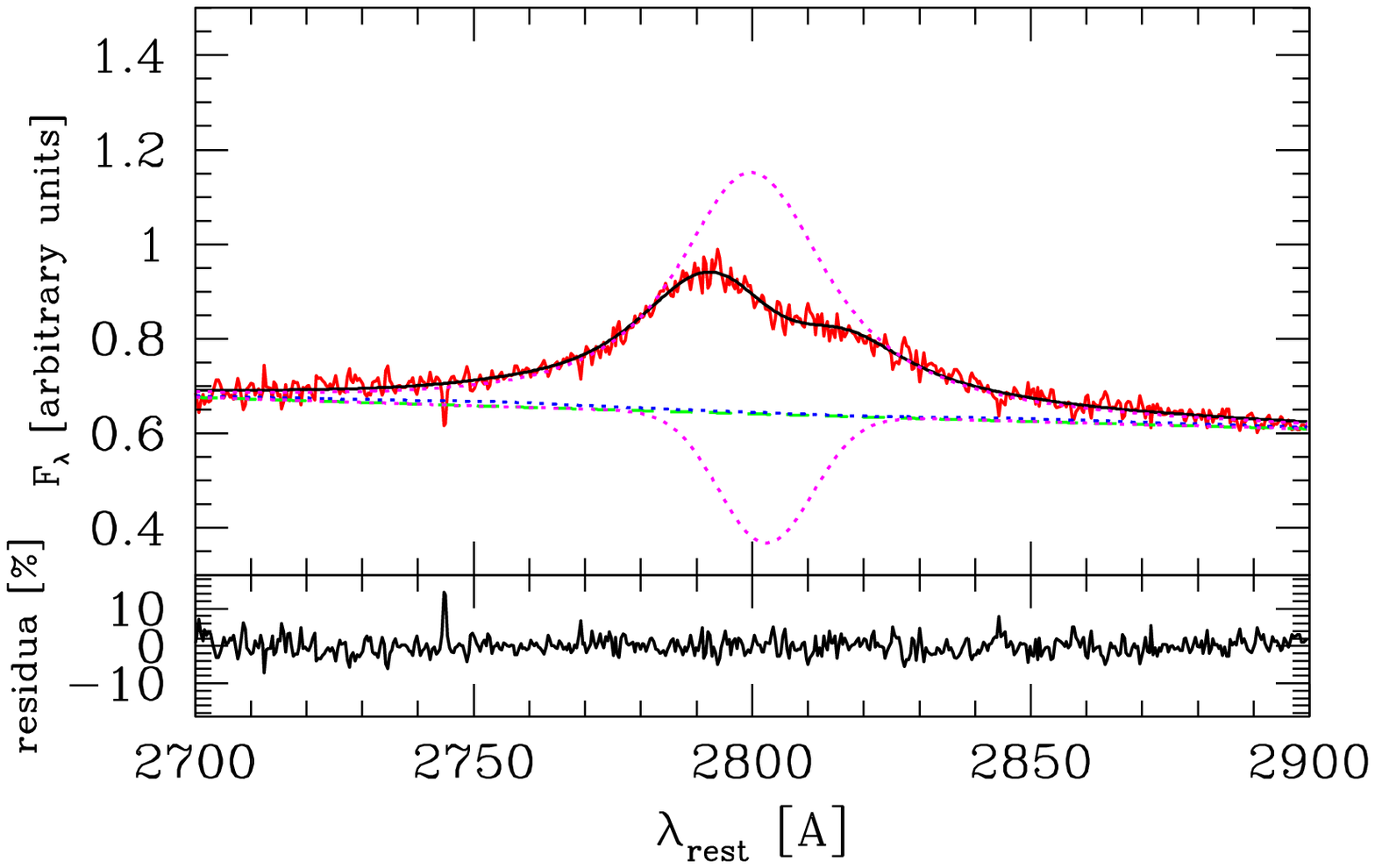}
\vskip - 4.0 true cm
  \includegraphics[width=0.9\hsize]{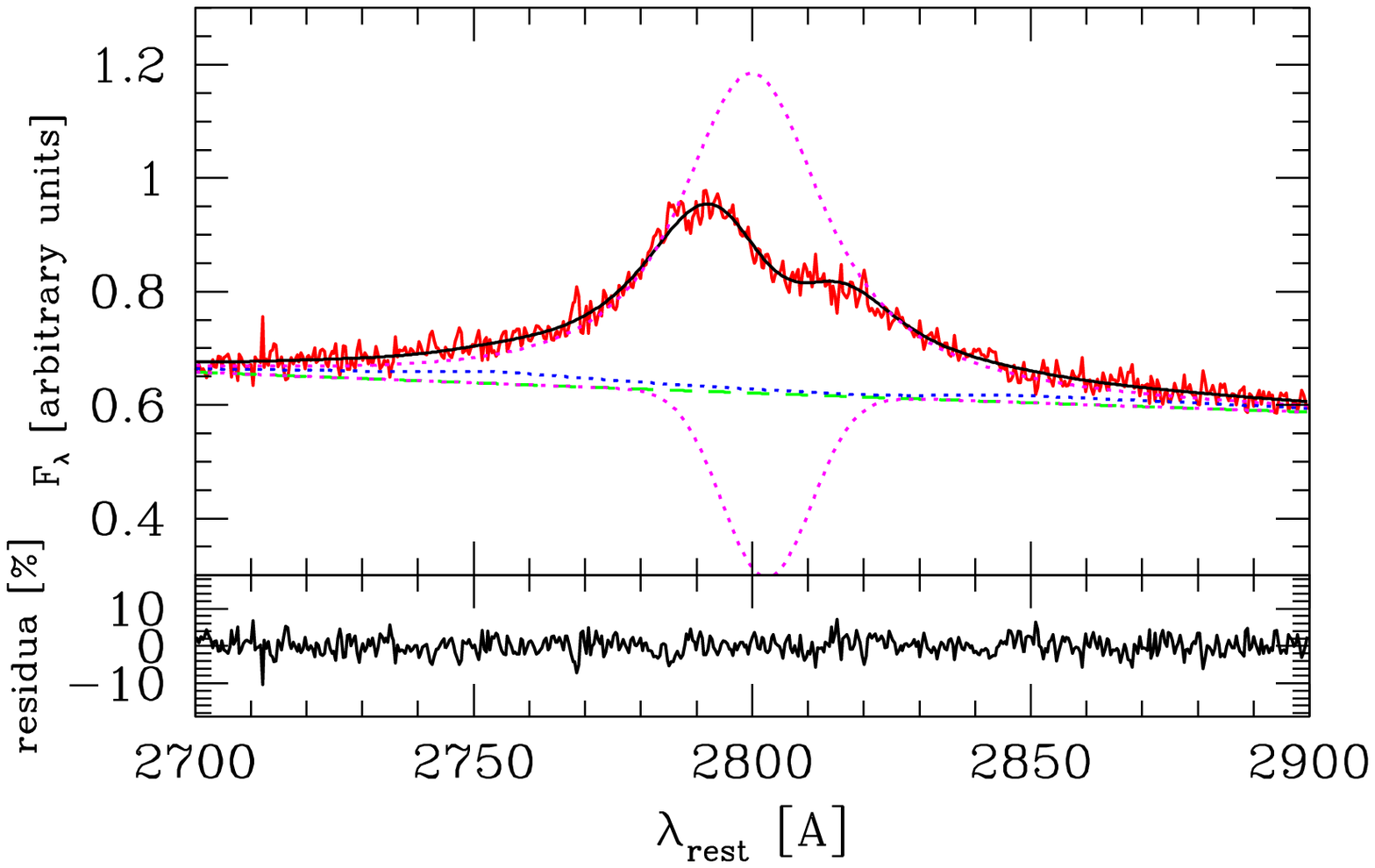}
\vskip - 4.0 true cm
   \includegraphics[width=0.9\hsize]{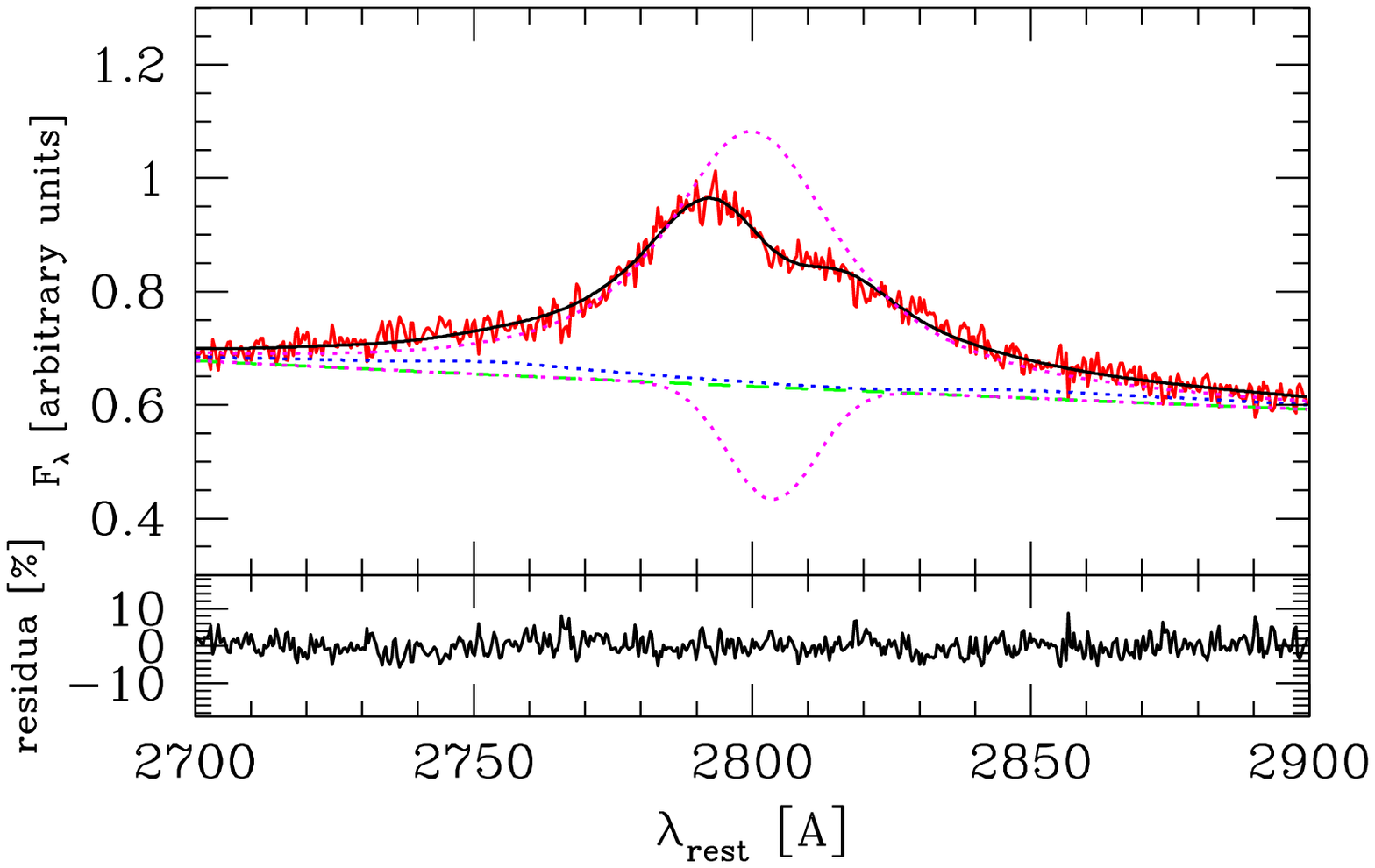}
\vskip - 4.0 true cm
  \includegraphics[width=0.9\hsize]{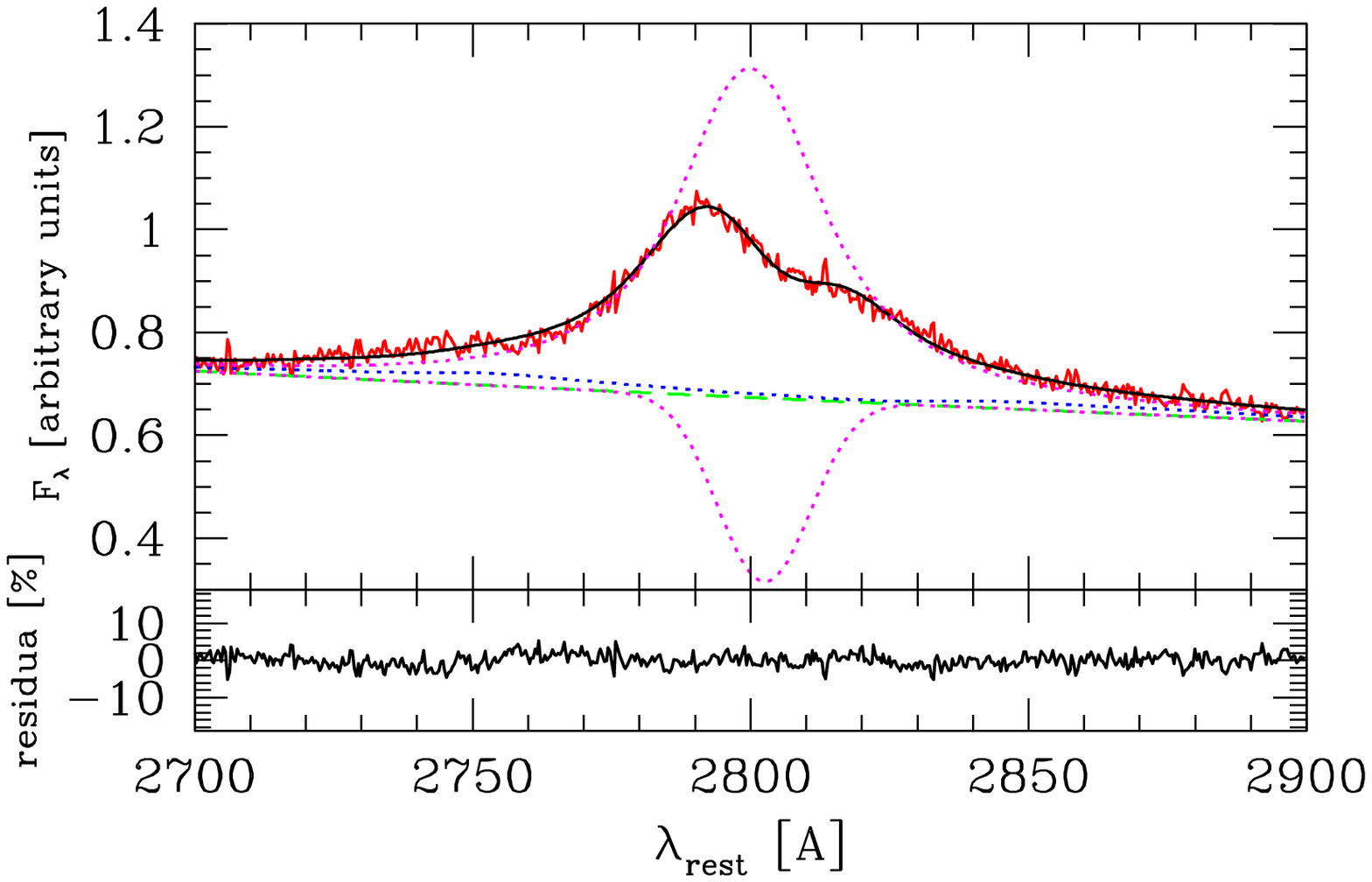}
\vskip - 0.5 true cm
   \caption{The best fit and residuals for five observations, model M (see Table~\ref{tab:absorpcja} for model parameters); continuous lines show the model and the data, dashed lines give the underlying power law,  the dotted blue line represents the kinematically blurred Fe II pseudo-continuum, and dotted magenta lines mark the emission and absorption components of the Mg II.}
              \label{fig:widma_absorpcja}%
    \end{figure}

Since we do not see significant difference in time behaviour of the two emission components we consider an alternative model: a single broad component
in emission and a narrower component in absorption. Such a model would immediately explain why the two wings, or emission components, vary together. We explored this interpretation in detail only for Fe template 13.  We modeled emission as a single Lorentzian, and absorption as a single Gaussian. Such a model has exactly the same number of parameters as the previous model with two emission components.  The resulting best fits for all five observations are shown in Fig.~\ref{fig:widma_absorpcja}.  For observation 2 the fit with absorption is better than for both components in emission, for observations 1 the two models are equally good, for observation 3, 4 and 5 the pure emission model is better by 7, above 100 and 73 in $\chi^2$. Thus, statistically, the absorption model is not favored. The errors of the fitted parameters are large since the absorption and emission have to compensate at $\sim ~ 2800$ \AA. 

\begin{table*}
\caption{Spectral fits for the template 13 of Fe II,  a Lorentzian component in emission, and a Gaussian component in absorption, for five individual spectra obtained with SALT between December 2012 and March 2014.}   
\label{tab:absorpcja}      
\centering                          
\begin{tabular}{l r r r r r r r}        
\hline\hline      
Obs.          & slope & Mg II  & Mg II       & Mg II       & Mg II       & Fe II  &$\chi^2$  \\
              &       & EW     & FWHM        & EW          & FWHM          & EW     &          \\
              &       & emission & emission  & absorption & absorption &    &          \\
              &        & \AA    & km s$^{-1}$  & \AA         &   km s$^{-1}$         & \AA    &          \\
\hline                        
1             &   -1.501       & 36.62 &  3109  &  -16.18     &   1618       &  1.57       &  269.3      \\   
2             &   -1.458       & 35.60 &  3175  &  -14.18     &   1577       &  1.63       &  401.3      \\   
3             &   -1.587       & 38.81 &  3000  &  -16.49     &   1452       &  3.49       &  207.6      \\   
4             &   -1.893       & 36.70 &  3809  &   -9.62     &   1455       & 3.78        &  722.6      \\   
5             &   -2.021       & 41.38 &  3072  &  -17.61     &   1561       & 3.71        &  300.9      \\   

\hline                                   

\end{tabular}
\end{table*}

\subsection{Rms spectra}

It is frequently advocated (e.g. Peterson et al. 2004) that rms spectra are better for the black hole mass determination. The shape of the rms spectrum may also indicate whether the two-component emission model is more justified, since in that case the variability of the two components may be different. We thus attempted to obtain such a spectrum for CTS C30.10 at the basis of four SALT spectra. The spectra, corrected with the use of the spectroscopic standard are not yet properly normalized. In order to obtain the correct normalization we use the photometric results. Three of the spectroscopic observations overlap with the OGLE photometry, observation 3 is in the gap. However, during this period the source seems to be systematically lowering its luminosity so we used the interpolation between the available photometric points. 
After this renormalization, we can obtain the rms spectra with the usual definition (see e.g. Peterson et al. 2004)
\begin{equation}
F_{rms}(\lambda) = \bigl[{1 \over (N - 1)} \sum\limits_{i=1}^N (F_i(\lambda) - \overline{F(\lambda)})^2\bigr]^{1/2},
\end{equation}
where $ \overline{F(\lambda)}$ is the average value obtained from a set of $N$ spectra.

\begin{figure}
   \centering
   \includegraphics[width=0.8\hsize]{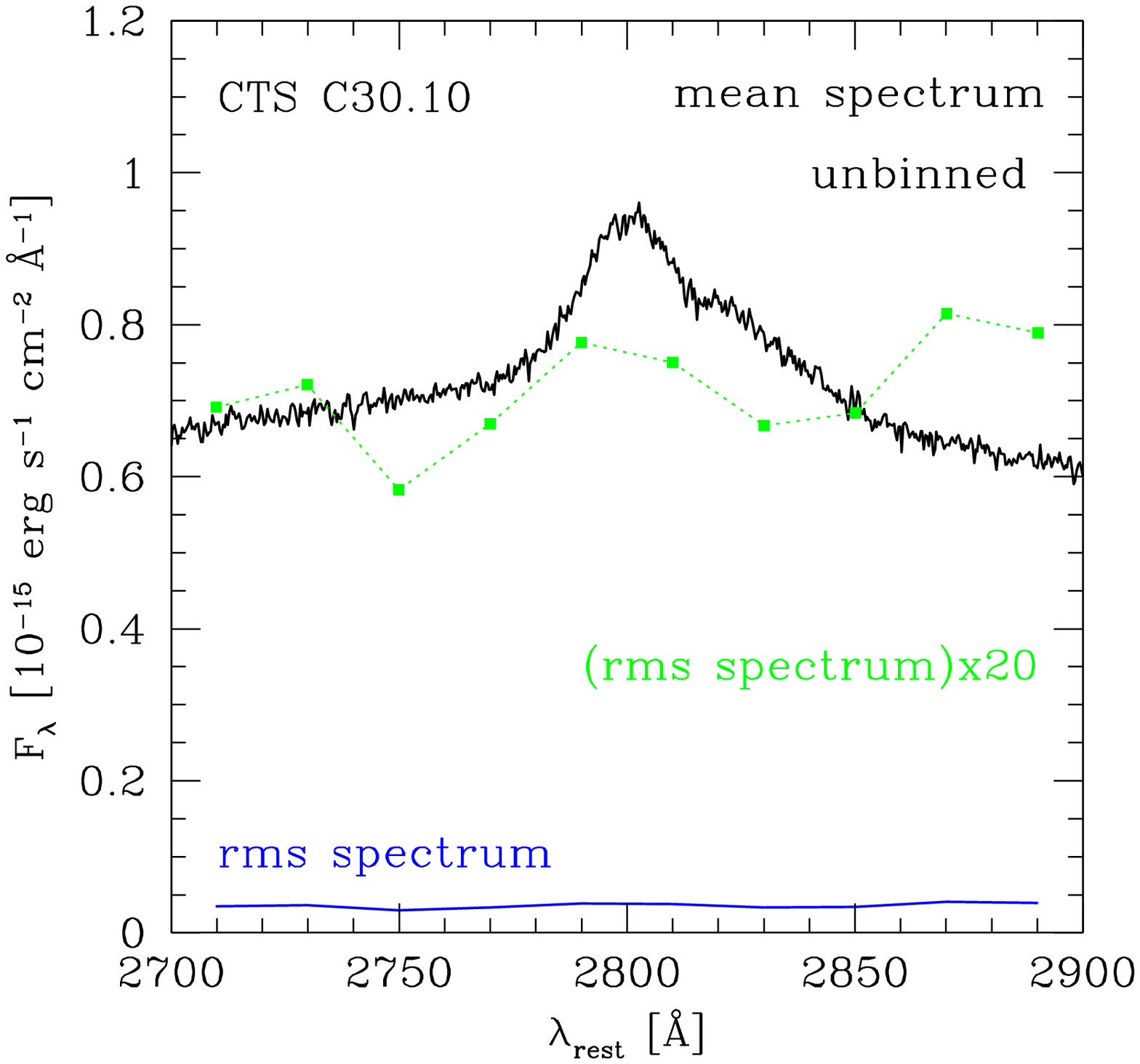}
   \caption{The mean (black line) and the binned (blue line) rms spectrum of CTS C30.10. We also plot the rms value multiplied arbitrarily by a factor 20 (green line) for a better visibility.}
              \label{fig:rms_spectrum}%
    \end{figure}

The resulting spectrum is shown in Fig.~\ref{fig:rms_spectrum}. The quality of the rms spectrum is very low since quasars vary slowly and much longer time separations are needed to measure the spectrum variations more accurately. We had to bin the spectra considerably in order to get any signal. The variability seems lower in the regions contaminated more strongly by Fe II while there is no clear difference in variability between the two kinematic components of Mg II line.
We tried to compare the variability of the two components by using bin size covering the two components in an optimum way (2770 - 2815 \AA ~and 2815 = 2860 \AA) but the ratio of the $F_{var}/F_{mean}$ in such bins was quite similar, $0.037 \pm 0.002$ and $0.036 \pm 0.002$. 
 
%
%

\subsection{Global parameters}

The Mg II line in CST C30.10 shows two kinematic components and it is not obvious a priori how to apply the usual methods of 
the black hole mass determination. We therefore considered four possibilities: (i) to use the component 1, (ii) to use the component 2, (iii)
to use a single-component fit to the spectrum, and (iv) to use the emission line parameters from the emission/absorption model, although the fit quality is much worse in two last cases. We used three formulae 
for the black hole mass determination. The first one comes from Kong et al. (2006), their Eq. 7,  the second one comes 
from Wang et al. (2009), their Eq. (10), and the third one from Trakhtenbrot \& Netzer (2012). This last paper is particularly relevant since it contains the correction for the doublet character of Mg II, important for lines with the FWHM below 4000 km s$^{-1}$, and included in our analysis.
The formulae are based on the knowledge of the monochromatic luminosity at 3000 \AA~ and the FWHM of Mg II line but differ 
slightly in the values of the coefficients. The monochromatic flux at 3000 \AA~ was obtained from the photometric flux in V band
(i.e. 2840 \AA~ rest frame) using the standard cosmology ($H_o = 71$ km s$^{-1}$ Mpc$^{-1}$, 
$\Omega_m = 0.270$, $\Omega_{\Lambda} = 0.730$). It was then extrapolated to the 3000 \AA~ using the fitted spectral slope.

We also calculated the corresponding Eddington ratios. Bolometric luminosity was obtained from the monochromatic flux at 3000 \AA~
applying the bolometric correction from Richards et al. (2006), $L_{bol} = 5.62 \times \lambda L_{\lambda}(3000 \AA)$.  

\begin{table*}
\caption{Global parameters of CTS C30.10 from Observation 2}   
\label{tab:glob}      
\centering                          
\begin{tabular}{l r r r r r r}        
\hline\hline      
Mg II line shape     &  BH mass	&  $L_{bol}/L_{Edd}$	&  BH mass	&   $L_{bol}/L_{Edd}$	& BH mass 	&   $L_{bol}/L_{Edd}$ \\
			    &	Kong	& Kong			&  Wang		&	Wang		& Trakhtenbrot	&	Trakhtenbrot	\\
component 1 only     & $6.8 \times 10^8 M_{\odot}$    &    1.6               &   $1.0 \times 10^9 M_{\odot}$      &  1.1          &   $1.4 \times 10^9 M_{\odot}$           &   0.8\\
component 2 only     & $9.9 \times 10^8 M_{\odot}$    &    1.1               &   $1.4 \times 10^9 M_{\odot}$      &  0.8          &   $2.0 \times 10^9 M_{\odot}$           &   0.5\\
single component fit & $2.4 \times 10^9 M_{\odot}$    &    0.6               &   $2.7 \times 10^9 M_{\odot}$      &  0.4          &   $4.9 \times 10^9 M_{\odot}$           &   0.2\\
absorption model     & $8.5 \times 10^8 M_{\odot}$    &    1.3               &   $1.2 \times 10^9 M_{\odot}$      &  0.9          &   $1.7 \times 10^9 M_{\odot}$           &   0.6\\  
\hline                                   
\end{tabular}
\end{table*}

 
The results obtained for the Observation 2 are given in Table~\ref{tab:glob}. The values based on the component 1 indicate rather high Eddington ratio, for all the formulae. This is rather unexpected since the sources with two-component and generally broader lines are believed to belong to the population B sources with lower Eddington ratio (Marziani et al. 2013ab).  The Eddington ratio for the type A quasar LBQS 2113-4538 studied before was 0.7 at the basis of the same, Trakhtenbrot \& Netzer (2012) formulae.  If the classification into A and B sources is applied only to the component  1, the source should belong to class A. On the other hand, Marziani et al. (2013b) notice that some quasars do not quite follow the classification and the BC class is actually a combination of type A2 and type B1 sources. The results based on absorption model give also rather high Eddington ratio. Although the red and blue wings in this case are fitted with a single component, line maximum is much higher than observed (peak is hidden by absorption). Component 2 give lower Eddington ratio but the fact that this component is not accompanied by the usual Fe II suggests rather different origin than from the clouds close to the disk and in Keplerian motion. 

Single-component fit implies moderate Eddington ratio and in the low quality data such a solution would be satisfactory. However, we cannot neglect the fact that a single component fit does not represent well the data.
Results based on other observations are rather similar, since neither the monochromatic luminosity nor the line kinematic width vary
significantly. 

We checked whether part of the problem lies in partial contamination of the V band by the Mg II and Fe II emission. For that purpose, we calculated the fraction of the Mg II and Fe II emission in the V filter, taking into account the standard V filter profile and the spectral decomposition for Observation 2, model M. The contamination is not strong - Mg II contributes 5.4 \%, and Fe II  only 2.9 \%. This does not change the obtained Eddington ratio considerably, reducing it from 0.77 down to 0.71, for Trakhtenbrot \& Netzer (2012) formulae.

As an independent check of the appropriate values of the black hole mass and accretion rate in our object, we performed simple fits of an accretion disk to the broad band spectrum of CTS C30.10.

We tested the black hole mass values based on the Mg II width, as well as checked an applicable parameter range. As a disk model, we applied the simplest Novikov-Thorne model, with relativistic effects inluded, but without any color corrections or limb darkening effects to the local spectra.  

\begin{figure}
   \centering
   \includegraphics[width=0.8\hsize]{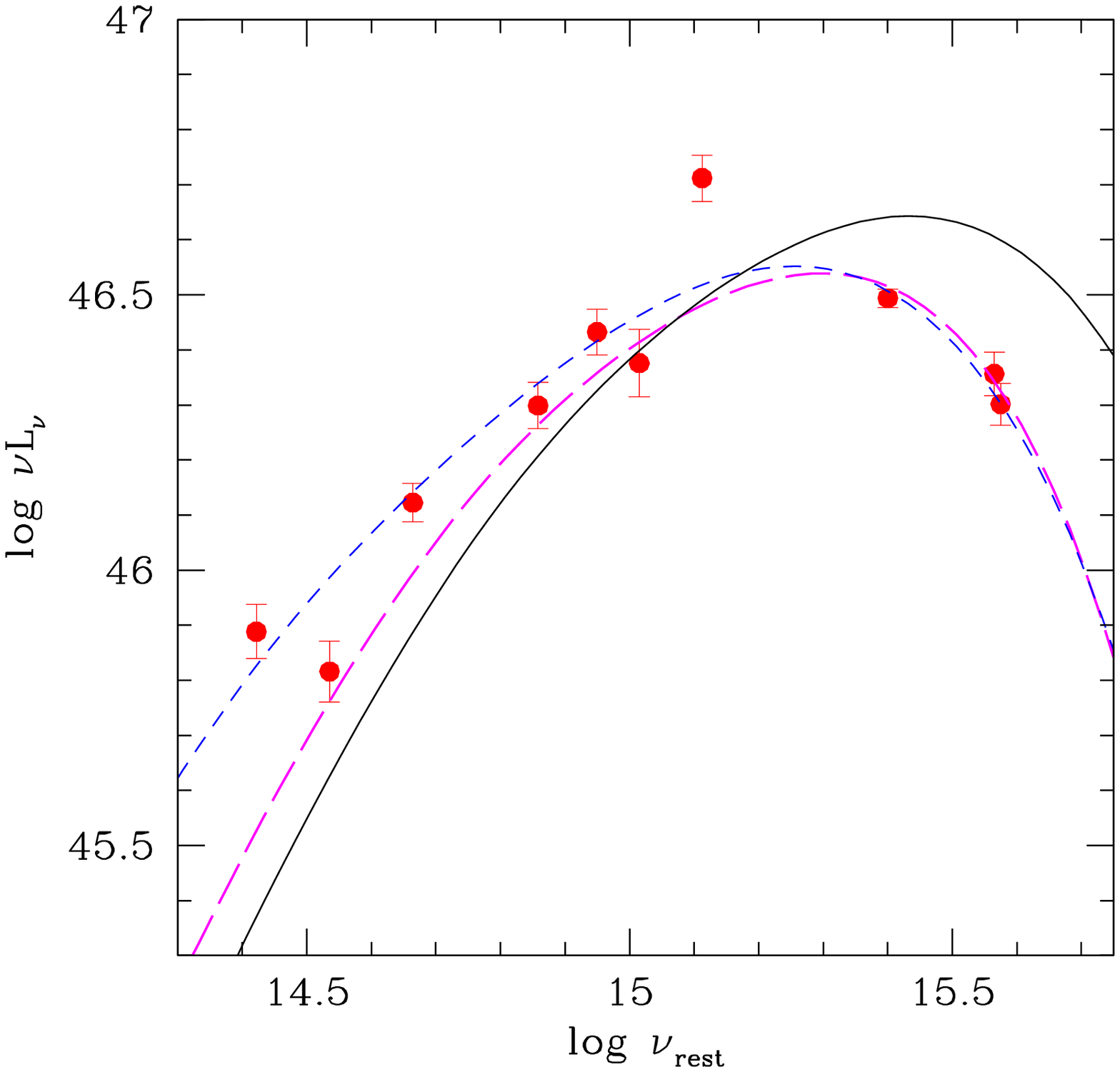}
   \caption{The fits of Novikov-Thorne model to CTS C30.10 in the optical/UV band: (i) the black hole mass and Eddington ration taken from Table~\ref{tab:glob}, component 1 only, Thrakhtenbrot formula (continuous line; Schwarzschild solution, best fit inclination of 63 deg), (ii) the same black hole mass but accretion rate arbitrarily fixed $L/L_{Edd} = 0.35$ (long dashed line; best fit inclination - top view), (ii) single component fit (short dashed line; black hole spin $a = 0.9$, best fit inclination of 41 deg}
              \label{fig:NT_fits}%
    \end{figure}

The values of the black hole mass and the Eddington ratio derived from the component 1 and Trakhtenbrot \& Netzer (2012) formulae, even for the Schwarschild black hole, overpredict the observed points above 3000 \AA~ (see Fig.~\ref{fig:NT_fits}). Any positive black hole spin would make the fit only worse, strong counter-rotation somewhat eases the problem. In addition, the implied inclination angle is quite large - otherwise the discrepancy above 3000 \AA~ is even stronger. 
If we use the black hole mass and the Eddington ratio from a single component fit, the predicted maximum temperature in the disk drops significantly and the
Schwarschild model represents the overall spectrum reasonably well. The required inclination angle is again rather large (60 deg) for a type 1
object. However, if we take the black hole mass derived from the component 1 and we simply assume the Eddington ratio much lower than implied by the standard formula we decrease the inclination angle easily. For the top view the implied Eddington ratio, 0.35, is relatively low and more consistent with 
the object line properties. Higher values of the black hole spin require to adopt higher value of the black hole mass than given by component 1. For example, black hole spin 0.9, black hole mass from a single component fit ($4.9 \times 10^9 M_{\odot}$) gives a very nice fit for an accretion rate of 0.055 and inclination angle of 41 deg. Formally, the last solution shows lower $\chi^2$ than solutions for a Schwarzschild black hole but we do not think that the quality of the available broad band spectrum is good enough to relay on the conclusions based on the curvature of the high frequency part of the spectrum which differentiates between the various spin models as a result of the relativistic effects. However, the low value of the accretion rate is another argument in favor of this class of solutions.   

%
%

\subsection{Covering factor}

In order to have an additional possible insight into the geometry of the emitting region we estimated the covering factor required to produce the observed Mg II line equivalent width. We used the observed continuum but in order to have rather smooth input continuum we shifted down the B-band outlying point from USNO by 0.22 dex down in order to have more realistic SED. This accounts for Fe II contamination (7 \%) and variability amplitude seen in the CATALINA data.
Calculations were performed with version 13.02 of Cloudy (Ferland et al. 2013). We adopted a single cloud approach since we do not have other emission lines. We assumed the hydrogen column of $10^{23}$ cm$^{-2}$ and a range of local densities between $10^{12}$ cm$^{-3}$ and $10^{14}$ cm$^{-3}$, appropriate for a LIL part of the BLR and consistent with the local densities at the accretion disk surface at the BLR distances (e.g. Rozanska et al. 2014). The results are not unique, mostly because of the uncertainty of the continuum in the E-UV and soft X-ray band. If we assume the broad band continuum as displayed in Fig.~\ref{fig:broad_band} we obtain the covering fraction in the range of 0.51 - 0.83. It is very high, but the covering factor is an old  problem of the BLR, and various ways has been postulated to solve it (see e.g. MacAlpine et al. 2003). However, if we assume that the source has significant soft X-ray excess and instead of using a photon index $\Gamma = 2.0$ in the X-ray band we simply interpolate between the last UV point and X-ray point (this gives $\Gamma = 2.65$) the covering factor reduced to 0.24 - 0.39.  


\subsection{Prospects for time delay measurement}

Our measurements of the line intensity cover 15 months, and the same period is covered by the OGLE precise photometry. Earlier photometric monitoring is available from CATALINA survey. We show these data in  Fig.~\ref{fig:lightcurve}. Mg II line intensity is based on the two emission component of Mg II (see Sect.~\ref{sect:two_emiss}). Both photometry data sets confirm the presence of clear long lasting trends in continuum which should allow for the time delay measurement between the line and continuum. The line intensity does not vary significantly during the first four observations but we see a drop in the Observation 5 which may indicate that the line does respond to the change of the continuum. In addition, the CATALINA photometry is of too low quality to attempt any delay measurement at this point. Thus so far the distant quasar monitoring has brought only the tentative measuremnt of a single object by Kaspi et al. (2007).

We can use the source redshift and the extinction-corrected interpolated monochromatic luminosity of $1.44 \times 10^{46}$ erg s$^{-1}$ at 5100 \AA ~in the rest frame to estimate the expected time delay between the line and continuum. Using either the observational relation between the delay and flux from Bentz et al. (2013)
\begin{equation}
\log R_{BLR} = 1.555  + 0.542  \log {\lambda L_{\lambda}^{5100 \AA} \over 10^{44} {\rm erg~ s}^{-1}} {\rm  [lt. ~days]}
\end{equation}
obtained for nearby objects or applying the dust-origin formula to the BLR size
\begin{eqnarray}
\log R_{BLR} = 1.244  + 0.5  \log {\lambda L_{\lambda}^{5100 \AA} \over 10^{44} {\rm erg~ s}^{-1}} \nonumber\\
+ \log (1 + \sin i) - 0.5 \log \cos i  {\rm  [lt. ~days]}
\end{eqnarray} 
give 530 days and 395 days correspondingly, in the rest frame. In this last case, we assumed an inclination angle $i$ of the disk equal 40 deg and the dust temperature of 1000 K in the Czerny \& Hryniewicz (2011) formulae. Thus much longer monitoring is clearly needed.


   \begin{figure}
   \centering

\includegraphics[width=0.99\hsize]{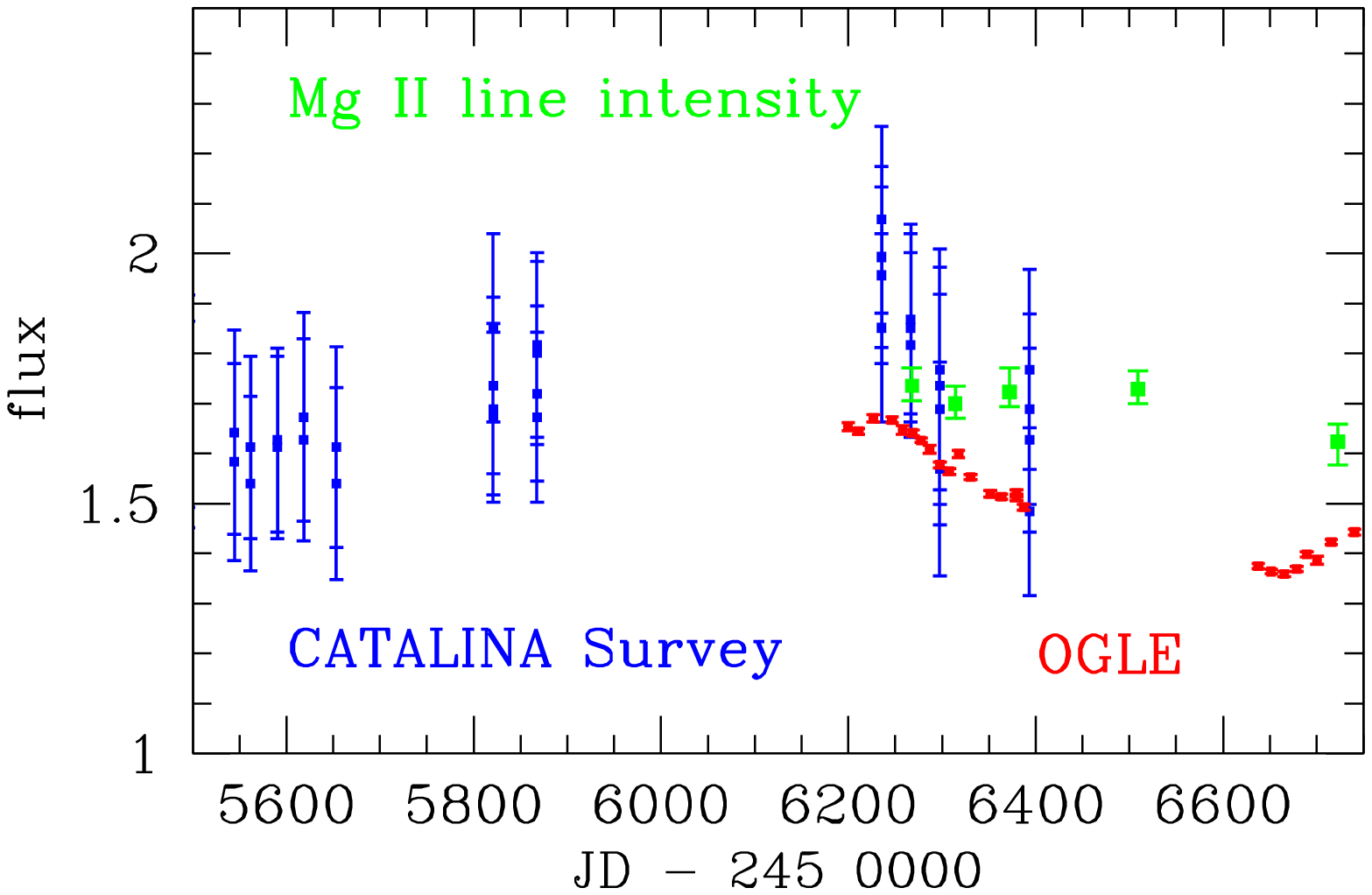}
   \caption{The time evolution of the V-band flux and the Mg II line intensity. The continuum is in units of $3.77  \times 10^{-16}$ erg s$^{-1}$ cm$^{-2}$, and the line flux was scaled by a factor of $1.7 \times 10^{-14}$ erg s$^{-1}$ cm$^{-2}$. }
              \label{fig:delay}%
    \end{figure}
%
%



\section{Discussion}

We analyzed in detail five long-slit spectroscopic observations of the quasar CTS C30.10 with the aim to determine the shape and the variability of the Mg II broad emission line and the Fe II pseudo-continuum. The object is representative for
quasars showing complex, two-component shape of Mg II, and thus differ qualitatively from single-component quasars like LBQS 2113-4538 (Hryniewicz et al. 2014). According to Sulentic et al. (2007) classification, CTS C30.10 thus belongs to type B quasars.

\subsection{Line properties and quasar variability}

Our spectroscopy covers 15 months, and during this period the source systematically gets fainter, accoring to OGLE photometry. The source changes the luminosity by 0.23 mag, i.e. it is more variable than the type A quasar studied by us before (amplitude of 0.05 mag in 6 months). This is consistent with spectral classification as type B source, since in the optical/UV band the luminosity decreases with the Eddington ration (e.g. Zuo et al. 2012; Ai et al. 2013). 

Our SALT spectroscopic data show that all spectral components do vary. The continuum gets bluer with the decrease of the source luminosity. However, this trend may be an artifact of the behaviour of the Balmer continuum which was not modeled as a separate component. It is known that the Balmer continuum in general strongly contributes to the continuum variability at 3000 \AA~ (e.g. Kokubo et al. 2014). 
The total EW of the Mg II line systematically rises. This rise is mostly caused by the change of the continuum. The total line luminosity remains constant during the first four observations, at the mean level of $(2.93 \pm 0.03) \times 10^{-14}$ erg s$^{-1}$ cm$^{-2}$  it drops significantly down to  $(2.76 \pm 0.07) \times 10^{-14}$ erg s$^{-1}$ cm$^{-2}$ in the last observation. Thus the change in the line luminosity - only by $~\sim 6$ \% is much lower than the change in the continuum ($~\sim 24$ \%). This is roughly consistent with the Mg II variability level of Mg II line 5 times lower than the variability of the continuum, found by Kokubo et al. (2014) in SDSS data. The source thus seem typical type B quasar, well chosen for the monitoring studies.

The total EW of the Mg II line does not depend strongly whether it is modeled as a single kinematic component or two separate components. However, the two-component fits is always better than a single-component fit. The second (red) kinematic component is only slightly broader than the first (blue) component, and the relative shift is the same in all five data sets ($\sim 2780 $ km s$^{-1}$). Only the first kinematic component of Mg II line has the underlying Fe II contribution. We do not detect any contribution from the NLR, with an upper limit of 2 \%.  We do not detect yet any significant difference between the variability of the two components. Both the direct fitting and the rms study do not point towards a different time behaviour of the two kinematic components of the Mg II line. 

The EW of the Fe II pseudo-continuum measured in the 2700-2900 \AA~ band changes  monotonically, but the line intensity is also not monotonic as in the case of the Mg II. The accuracy of its measurement is lower then the Mg II measurement due to the coupling with the  slope of the continuum in data fitting. We do not have the reliable broad band spectrum to measure the continuum slope in the wavelength bands relatively free of the Fe II contamination.

Out of the several templates used, the template d12-m20-20-5 of Bruhweiler \& Verner (2008), convolved with a Gaussian of dispersion 900 km s$^{-1}$ (model M in Table~\ref{tab:wszystkie}) is overall the best but several other theoretical templates are not much worse. This template assumes an order of magnitude higher density $10^{12}$ cm$^{-3}$ than the best template for type A objects like LBQS 2113-4538 and I Zw 1 (Hryniewicz et al. 2014; Bruhweiler \& Verver 2008), with the two other parameters (turbulent velocity and ionization flux) being the same.

\subsection{Black hole mass and the broad band spectrum}

Two-component character of the Mg II line creates a problem for the black hole mass determination. Our lack of a complete picture of the BLR dynamics leads to a need of some arbitrary choice whether to use the whole line profile and its FWHM, or one of the kinematic components. We did the broad band spectra fitting for various choice of the formulae and components (see Table~\ref{tab:glob}). Overall, the mass determination based on the component 1 is in reasonable agreement with fitting the Novikov-Thorne model to the broad band spectra ($1.4 \times 10^9 M_{\odot}$ for a non-rotating black hole) but the bolometric luminosity seems to be considerably over-predicted by the standard formulae. Disk fitting indicates lower inclination and lower Eddington ratio (only 0.35, more consistently with type B source). The disk fits are not unique, and the black hole mass up to $2.0 \times 10^9 M_{\odot}$ is equally good but requires higher Eddington ratio (up to 0.5) and higher inclination (60 deg). Lower inclination might also explain why the FWHM of the component 1 is relatively low, only $\sim 2800$ km s$^{-1}$. The overall width is much larger, due to the presence of the second kinematic component. If we allow for a large spin of the black hole, then the black hole mass value based on a single component fit is far more appropriate, and the Eddington ratio in such a fit is much lower. This might favor using the whole line profile. On the other hand, modeling of the quasar spectra in the far UV is problematic in most of quasars (see e.g. Lawrence 2012; but see Czerny et al. 2011 for an example of the successful spin determination based on broad band fitting of SDSS J094533.99+100950.1). Most of the objects show a turn-off at $\sim 1000$ \AA~ in the rest frame, corresponding to the maximum of the disk temperature of about 50 000 K (Laor \& Davis 2014), and this might result from the vigorous line-driven wind outflow from the disk surface at higher temperatures.

\subsection{Two-component BLR region for LIL}

Type A quasars show a single component Mg II line of a Lorentzian shape (e.g. Laor et al. 1997b; Veron-Cetty et al. 2001, Sulentic et al. 2002, Zamfir at el. 2010;  Sulentic et al. 2009, 2011; Shapovalova et al. 2012), but in type B sources, having lower Eddington ratio, a second kinematic component appears.
Its presence remains unexplained. It might imply the presence of another emitting region, or some further reprocessing (absorption or scattering) in the circumnuclear region. It might also be the direct signature of the emission coming from an accretion disk which is expected to have double-peak profile.

The High Ionization Lines (HIL) show the assymetry and systematic shifts with respect to the NLR which is naturally interpreted as the result of emission from an outflowing wind. LIL lines usually do not show such strong shift but the assymetry clearly appears in some sources, like the one studied here, and a need for a multi-component fit appears. This is frequently seen in the H$\beta$ structure (e.g. Hu et al. 2012) since this line is far better studied than Mg II but the overall trend is similar in both lines (e.g. Marziani et al. 2013b). In the case of our source we cannot determine the quasar rest frame. We have searched for weak narrow emission lines in the available spectrum but we were unable to identify any of them in a reliable way. 

It is interesting to note that radio-loud sources also can display line profiles of the type B. In particular, the Mg II shape of 3C 279 also shows a second, redshifted component, with the velocity shift between the two being of order of 4000 km S$^{-1}$ (Punsly 2013). Since 3C 279 is a blazar observed at very low inclination angle (below $\sim 5$ deg; e.g. Bloom et al. 2013), Punsly (2013) suggests that the red part of Mg II comes from upscattering by the plasma either infalling close to the jest axis on the observer's side, or outflowing on the counter-jet side. The two components in 3C 279 varied in a different way so both possibilities are equally likely. However, in our case the two components seem to vary in the same way, i.e. the inflow interpretation offers the most probable interpretation. 
Such a quasi-radial inflow close to the symmetry axis is predicted by the simulations of accretion onto black hole when the accreting material has a range of angular momentum (e.g. Proga \& Begelmann 2003). The inflow velocity is of the order of the local escape velocity, so if the region is radially somewhat more distant that the production region of Mg II, the requested redshift-inducing
speed of 2780 km s$^{-1}$  can be achieved. However, in this case the angular extension of the accreting region is large, of order of 30 deg around the symmetry axis so the significant velocity gradient would be expected along the line of sight to the observer. On the other hand, jet formation is accompanied by the cocoon developments, and for some parameters the backflowing plasma forms a relatively narrow elongated structure along the jet (e.g.  Massaglia et al 1996).  Such a structure might be a more attractive candidate for the scattering inflowing medium but in this case the scattering probability becomes too low for obvious geometrical reasons.  

On the other hand, long monitoring H$\beta$ line shape of NGC 5548 and 3C 390.3 (which belong to B class) shows that the two components actually come from a disk, the relative strength of the two slowly varies with time over the years, and the changes are well modeled by some kind of spiral structure within an accretion disk (see Shapovalova et al. 2009 and the references therein). The separation between the two peaks in NGC 5548 is about 3000 km s$^{-1}$, quite similar to that observed in our source. In other objects, like NGC 4151, the situation is more complex and multi-component origin of H$\beta$ is requested. 

\subsection{Cosmological application of type B objects at intermediate redshifts}

Reverberation studies of quasars can be used as new cosmology probes of the expansion of the Universe (Watson et al. 2011, Czerny et al. 2013; but see also 
Marziani \& Sulentic 2013, La Franca et al. 2014, Wang et al. 2013  for a different approach).
Optical observations of intermediate redshift quasars give opportunity to observe Mg II line. Selection of type A objects is an advantage since the shape of the Mg II line is very simple
in this class, the whole line is well represented with a single Lorentzian shape. The line symmetry supports the conclusion about the absence of an inflow/outflow so the motion of the emitting clouds is then likely to be consistent with the Keplerian motion. Type B objects have more complex Mg II line, so it is not a priori clear whether the response of the line
components are the same. On the other hand, the overall variability level of type B objects in the optical/UV band is higher (e.g., Papadakis et al. 2000; Klimek et al. 2004, Ai et al. 2013) which makes easier the determination of the time delay. However, the observed variability level of Mg II in CTS C30.10 is still much lower than the continuum variability which is consistent with the reprocessing scenario. In this case we can expect that the line luminosity indeed may reflect the changes in the continuum and the measurement of the delay is likely to be possible. In the case of the CIV monitoring done with HET telescope, the CIV line varied more (Kaspi et al. 2007), and the dominance of the intrinsic changes made the determination of the delay difficult, despite seven years of monitoring. 
The underlying Fe II pseudo-continuum complicates the analysis in both types of objects, but it seems that a good theoretical template exists (one of the templates from 
Bruhweiler \& Verner 2008) which can be used almost universally to model this contribution. The quality of the SALT spectroscopy is high enough to measure the EW of the Mg II with accuracy of $\sim 0.25$ \AA,~ despite the Fe II uncertainty. 





\begin{acknowledgements}
Part of this work was supported by the Polish grants Nr. 719/N-SALT/2010/0, UMO-2012/07/B/ST9/04425. The spectroscopic 
observations reported in this paper were obtained with the Southern African Large Telescope (SALT),  
proposals 2012-2-POL-003 and 2013-1-POL-RSA-002. JM, KH, BC, MK and AS
acknowledge the support by the Foundation for Polish Science through the
Master/Mistrz program 3/2012. K.H. also thanks the Scientific
Exchange Programme (Sciex) NMSch for the opportunity of working at ISDC. 
The financial assistance of the South African National Research Foundation (NRF) towards
this research is hereby acknowledged (MB).
The OGLE project has received funding from the European Research Council
under the European Community's Seventh Framework Programme
(FP7/2007-2013) / ERC grant agreement no. 246678 to AU.
The Fe II theoretical templates described in 
Bruhweiler \& Verner (2008) were downloaded from the
web page http://iacs.cua.edu/personnel/personal-verner-feii.cfm with the permission of the authors.
Part of this work is based on archival data, software or on-line services provided by the ASI Science Data Center (ASDC).
This research has made use of the USNOFS Image and Catalogue Archive
   operated by the United States Naval Observatory, Flagstaff Station
   (http://www.nofs.navy.mil/data/fchpix/).
This publication makes use of data products from the Wide-field Infrared Survey Explorer, which is a joint project of the University of
California, Los Angeles, and the Jet Propulsion Laboratory/California Institute of Technology, funded by the National Aeronautics and
Space Administration.

This publication makes use of data products from the Two Micron All Sky Survey, which is a joint project of the University of
Massachusetts and the Infrared Processing and Analysis Center/California Institute of Technology, funded by the National Aeronautics
and Space Administration and the National Science Foundation.
The CSS survey is funded by the National Aeronautics and Space
Administration under Grant No. NNG05GF22G issued through the Science
Mission Directorate Near-Earth Objects Observations Program.  The CRTS
survey is supported by the U.S.~National Science Foundation under
grants AST-0909182.
Based on observations made with the NASA Galaxy Evolution Explorer.
GALEX is operated for NASA by the California Institute of Technology under NASA contract NAS5-98034. 
This research has made use of the NASA/IPAC Extragalactic Database (NED) which is operated by the 
Jet Propulsion Laboratory, California Institute of Technology, under contract with the National Aeronautics and Space Administration. 
\end{acknowledgements}

\end{document}